\documentclass[aps,prd,superscriptaddress,amsmath,twocolumn,10pt]{revtex4}

\usepackage{color,hangcaption,hhline,psfrag,rotating,amssymb}
\usepackage[hang,nooneline]{subfigure}
\usepackage{dcolumn}
\usepackage{verbatim}
\usepackage{bm}

\begin{document}
\title{$V_{cs}$ from $D_s \rightarrow \phi \ell \nu$ semileptonic decay and full lattice QCD}

\author{G.~C.~Donald}
\thanks{Current address: School of Mathematics, Trinity College, Dublin 2, Ireland (donaldg@tcd.ie)}
\affiliation{SUPA, School of Physics and Astronomy, University of Glasgow, Glasgow, G12 8QQ, UK}
\author{C.~T.~H.~Davies}
\email[]{christine.davies@glasgow.ac.uk}
\affiliation{SUPA, School of Physics and Astronomy, University of Glasgow, Glasgow, G12 8QQ, UK}
\author{J.~Koponen}
\affiliation{SUPA, School of Physics and Astronomy, University of Glasgow, Glasgow, G12 8QQ, UK}
\author{G. P. Lepage}
\affiliation{Laboratory of Elementary-Particle Physics, Cornell University, Ithaca, New York 14853, USA}

\collaboration{HPQCD collaboration}
\homepage{http://www.physics.gla.ac.uk/HPQCD}
\noaffiliation

\date{\today}

\begin{abstract}
We determine the complete set of axial and vector form factors for 
the $D_s \rightarrow \phi \ell \nu$ decay from full lattice 
QCD for the first time. The valence quarks are implemented 
using the Highly Improved Staggered Quark action and 
we normalise the appropriate axial and vector currents 
fully nonperturbatively. The $q^2$ and angular distributions 
we obtain for the differential rate agree well with those 
from the BaBar experiment and, from the total branching 
fraction, we obtain $V_{cs} = 1.017(63)$, in good 
agreement with that from $D \rightarrow K \ell \nu$ semileptonic decay. 
We also find the mass and decay constant of 
the $\phi$ meson in good agreement with experiment, 
showing that its decay to $K\overline{K}$ (which we do 
not include here) has at most a small effect.  
We include an Appendix on nonperturbative renormalisation of the complete set of staggered vector and axial vector bilinears needed for this calculation.
\end{abstract}


\maketitle

\section{Introduction.} 
\label{sec:intro}
The analysis of weak semileptonic decays in 
which one meson changes into another 
and emits a $W$ boson provides a strong 
test of QCD. The test is complementary to that 
of comparing QCD predictions to experiment 
for the meson mass and leptonic decay constants, 
and in principle more stringent because, 
instead of just one number, the comparison involves 
the shape of a differential rate 
as a function of $q^2$, the square of the 
4-momentum transfer from initial to final meson. 
The QCD information that appears in the differential 
rate, and the functions of $q^2$ that are calculated in 
lattice QCD, are known as form factors.
Lattice QCD calculations have largely focussed 
on pseudoscalar to pseudoscalar decays where 
only one form factor contributes to the experimental rate. 
Accurate tests against experiment have been 
carried out for, for example, $D \rightarrow K \ell \nu$ decay~\cite{jonnadtok}. 
Here we study the pseudoscalar to vector 
decay, $D_s \rightarrow \phi \ell \nu$, in which 
3 form factors contribute to the experimental 
results. This allows us to compare angular 
distributions as well as differential rates 
in $q^2$, providing a more complete test of 
how QCD interactions that bind a quark inside
a meson affect the quark weak decay process. 
This is the first time this calculation has 
been done in full lattice QCD including the 
effect of sea quarks. 

The fundamental quark weak decay in 
$D_s \rightarrow \phi \ell \nu$ is a $c \rightarrow s$ transition and 
so comparison with experiment allows us to 
determine $V_{cs}$. This is then a
direct determination of this CKM element which 
is independent of other methods such as 
$D \rightarrow K \ell \nu$ semileptonic decay 
or $D_s$ leptonic annihilation. Although our 
result is currently not as accurate as these other methods, 
it nevertheless contributes 
to improving our confidence in the determination 
of $V_{cs}$ and the second row and column CKM 
unitarity tests in which it plays a key role. 
  
The $D_s \rightarrow \phi$ decay has initial and 
final mesons with no light valence quarks. This 
is useful for a lattice QCD calculation which includes 
light quarks with masses that are heavier than the 
physical values since 
it means that the extrapolation in 
the light quark mass to the physical point only affects 
sea quark contributions and so is relatively benign. 
The $\phi$ meson is likely to be more sensitive to 
light quark masses than the $D_s$ because 
it has a strong decay mode to $K\overline{K}$. 
The $\phi$ is below threshold for this decay in a lattice 
QCD calculation with heavier-than-physical light quark masses; 
it is only just above threshold when the light quarks 
have their physical masses. 
We will treat the $\phi$ as stable 
in our lattice QCD calculation. 
By comparing the $\phi$ decay constant we calculate 
on the lattice to the experimental rate, 
we can estimate the systematic error on matrix 
elements that can arise from ignoring the strong decay.

The $q^2$ range for $D_s \rightarrow \phi$ decay is 
not large, running from $q^2_{max} = (M_{D_s}-M_{\phi})^2$ 
$=0.898 ~\mathrm{GeV}^2$ to $q^2=0$. We can easily cover the entire 
range in a lattice QCD calculation, needing only 
$p_{\phi} = 0.719 ~\mathrm{GeV}$ in the $D_s$ rest frame 
to reach $q^2=0$. Discretisation errors are then 
small in a good discretisation such as the 
Highly Improved Staggered Quark formalism~\cite{HISQ_PRD} that 
we use here. Since the entire range in $q^2$ is 
covered  
we can make a detailed comparison to experimental 
distributions as a function of $q^2$ 
and we can integrate over $q^2$ to extract $V_{cs}$ from a comparison 
to experiment of the total branching fraction. 

The paper is laid out as follows: Section~\ref{sec:theory} 
describes the theoretical background and then Section~\ref{sec:lattice}
gives a general description of the lattice calculation. Section~\ref{sec:results} 
gives the details of the results, first for the $\phi$ meson and then 
for each of the form factors for $D_s \rightarrow \phi$ 
in turn, describing how they were 
calculated. A comparison to BaBar's experimental 
results is then made for the form factors and for the differential 
distributions as a function of $q^2$ and decay product angles and 
finally $V_{cs}$ is determined from the total rate. 
In Section~\ref{sec:discussion} we discuss the comparison 
between our form factors for $D_s \rightarrow \phi$ with those 
extracted by CLEO from experiment for $D \rightarrow K^*$. 
Section~\ref{sec:conclusions} gives our conclusions. 
In Appendix~\ref{sec:zfactors} we describe how to normalise 
all the form factors nonperturbatively and in Appendix~\ref{ptsplitmomentum}
we give more details for the specific case of the 1-link axial 
current operator.

\section{Theoretical background}
\label{sec:theory}
The matrix element of the hadronic weak $V-A$ current 
between the pseudoscalar $D_s$ and the vector $\phi$ meson
can be expressed in terms of form factors as~\cite{richman}
\begin{eqnarray}
&&\langle \phi(p^{\prime},\varepsilon) |V^{\mu}-A^{\mu} | D_s(p) \rangle  \label{eq:ffdef}\\
&=& \frac{2i\epsilon^{\mu\nu\alpha\beta}}{M_{D_s}+M_{\phi}}\varepsilon_{\nu}^*p^{\prime}_{\alpha}p_{\beta}V(q^2) - (M_{D_s}+M_{\phi})\varepsilon^{*\mu}A_1(q^2) \nonumber \\
&& +\frac{\varepsilon^*\cdot q}{M_{D_s}+M_{\phi}}(p+p^{\prime})^{\mu}A_2(q^2) + 2M_{\phi}\frac{\varepsilon^* \cdot q}{q^2} q^{\mu}A_3(q^2) \nonumber \\
&& -2M_{\phi}\frac{\varepsilon^* \cdot q}{q^2} q^{\mu}A_0(q^2) \nonumber .
\end{eqnarray}
Here $\varepsilon$ is the polarization vector of the 
$\phi$ meson and $q^{\mu}=p^{\mu}-p^{\prime \mu}$. The vector and axial vector currents 
are given in this case by $\overline{c} \gamma^{\mu} s$ and 
$\overline{c}\gamma^{\mu}\gamma^5 s$. 
$A_3$ is not an independent form factor since
\begin{equation}
A_3(q^2) = \frac{M_{D_s}+M_{\phi}}{2M_{\phi}}A_1(q^2) - \frac{M_{D_s}-M_{\phi}}{2M_{\phi}}A_2(q^2) .
\label{eq:a3def}
\end{equation}
We also have the kinematic constraint that $A_3(0)=A_0(0)$. 
The form factors that appear with factors of $q^{\mu}$ do 
not contribute significantly to the experimental rate when the $W$ boson 
decays to $e^+\nu_e$ or $\mu^+\nu_{\mu}$. The 
reason is that the 
expression in Eq.~\ref{eq:ffdef} is
dotted into the leptonic current, 
$L_{\mu}=u_{\overline{\ell}}\gamma_{\mu}(1-\gamma_5)u_{\nu}$, when forming the rate
and $q^{\mu}L_{\mu} \to 0$ as $m_{\ell} \rightarrow 0$. 
Thus the form factors that we need to calculate to compare to experiment
are $V(q^2)$, $A_1(q^2)$ and $A_2(q^2)$. 

In the lattice QCD calculation, to be described 
in section~\ref{sec:lattice}, all of the form 
factors will appear in the matrix elements of the 
vector and axial vector currents that we 
calculate (as in Eq.~\ref{eq:ffdef}), and we have to choose particular 
kinematic configurations to isolate each one. 
We will also use the matrix element of the pseudoscalar 
current, $P=\overline{c}\gamma_5 s$, to access some of the 
form factors. From the partially conserved axial current 
(PCAC), $\partial_{\mu}A^{\mu}=(m_1+m_2)P$,
 which is exact for staggered quarks we have  
\begin{equation}
\langle \phi(p^{\prime},\varepsilon) |P | D_s(p) \rangle = \frac{2M_{\phi}\varepsilon^*\cdot q}{(m_c+m_s)}A_0(q^2).
\label{eq:pme}
\end{equation}

\begin{figure}
\centering
\includegraphics[width=0.45\textwidth]{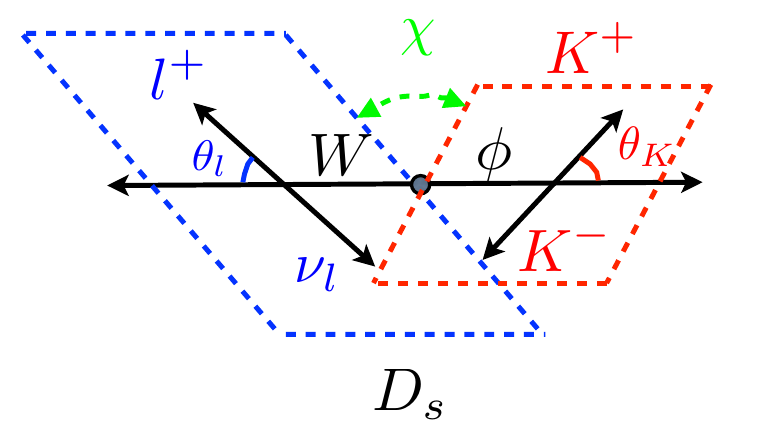}
\caption{Diagram to show the angles used for the differential rate for $D_s \rightarrow \phi \ell \nu$. $\ell \nu$ are drawn in the virtual $W^*$ rest frame and $K^+K^-$ in the $\phi$ 
rest frame. The angles are defined in the $D_s$ rest frame, however~\cite{richman}. }
\label{fig:angles}
\end{figure}

As well as comparing the shape of the extracted form 
factors to experiment we can also compare the 
differential cross-section in bins of $q^2$ or of the 
important angular variables for this decay. These 
angles are shown in Fig.~\ref{fig:angles} for results 
corresponding to the case where the $\phi$ is seen 
through its decay to $K^+K^-$. $\theta_{\ell}$ 
is the angle between the momentum of the charged 
lepton and that of the $W$ boson (= centre of momentum 
of the charged lepton and the neutrino) in the rest frame 
of the $D_s$. $\theta_K$ is 
the angle between the momentum of one of the $K$ mesons ($K^+$ for 
$D_s^+$ and $K^-$ for $D_s^-$) and the $\phi$ (= centre 
of momentum for both $K$ mesons). $\chi$ is the angle 
between the two planes, one defined by the $K$ meson pair and 
the other defined by the lepton pair. 

The differential rate for the decay is then given in terms of helicity 
amplitudes as~\cite{richman} 
\begin{eqnarray}
&&\frac{d\Gamma(D_s \rightarrow \phi \ell \nu, \phi \rightarrow K^+K^-)}{dq^2d\cos \theta_K d\cos \theta_{\ell} d\chi} = \label{eq:diffrate}\\
&& \frac{3}{8(4\pi)^4}G_F^2|V_{cs}|^2\frac{p_{\phi}q^2}{M^2_{D_s}}{\mathcal{B}}(\phi \rightarrow K^+K^-) \times \nonumber \\
&&\left\{ (1+ \cos \theta_{\ell})^2\sin^2 \theta_K |H_+(q^2)|^2 \right.\nonumber \\ 
&& + (1- \cos \theta_{\ell})^2\sin^2 \theta_K |H_-(q^2)|^2 \nonumber \\ 
&& + 4 \sin^2\theta_{\ell} \cos^2 \theta_K |H_0(q^2)|^2 \nonumber \\
&& + 4 \sin \theta_{\ell} (1+\cos\theta_{\ell})\sin \theta_K \cos \theta_K\cos \chi H_+(q^2)H_0(q^2) \nonumber \\
&& - 4 \sin \theta_{\ell} (1-\cos\theta_{\ell})\sin \theta_K \cos \theta_K\cos \chi H_-(q^2)H_0(q^2) \nonumber \\
&& \left. - 2 \sin^2 \theta_{\ell} \sin^2 \theta_K \cos 2 \chi H_+(q^2)H_-(q^2) \right\} . \nonumber 
\end{eqnarray}
$p_{\phi}$ is the momentum of the $\phi$ in the $D_s$ rest frame, in which we work. 
$H_{\pm}, H_0$ correspond to contributions from different $W$ helicities, and 
the $W$ and $\phi$ helicities are constrained to be the same because the parent 
meson has zero spin.  Helicity information on the quark 
produced in the weak decay is lost in a pseudoscalar to pseudoscalar transition, 
because the final meson has no helicity. Here, in a pseudoscalar to vector transition 
it is not lost, and thus the distributions give more information about the 
$V-A$ nature of the weak interaction. For a $c \rightarrow s$ decay we expect a 
predominantly $\lambda=-1/2$ $s$ quark to be produced, which can then form a 
helicity 0 or helicity -1 meson by combining with the spectator $\overline{s}$ 
to form a $\phi$. Thus we expect $H_-$ to dominate over $H_+$. 
The $W$ in a 
$c \rightarrow s$ decay is a $W^+$ and therefore decays to $\ell^+\nu$. A fast-moving 
$\ell^+$ will be predominantly $\lambda=+1/2$ and therefore preferentially 
thrown backwards in the $D_s$ rest frame to balance helicities. This explains 
the $\cos \theta_{\ell}$ distributions for the term proportional to $|H_-(q^2)|^2$~\cite{richman}.    
At low $q^2$, 
where the dominant configuration has the $\ell^+$ and $\nu$ in parallel, balancing 
the $\phi$, $H_0$ will dominate because the spins of $\ell$ and $\nu$ 
will cancel. 

The helicity functions are related to the form factors as 
\begin{equation}
H_{\pm}(q^2) = (M_{D_s}+M_{\phi})A_1(q^2) \mp \frac{2M_{D_s}p_{\phi}}{M_{D_s}+M_{\phi}}V(q^2)
\label{eq:hpm}
\end{equation}
and
\begin{eqnarray}
H_0(q^2) &=& \frac{1}{2M_{\phi}\sqrt{q^2}}\times \label{eq:h0}\\ 
&& [ (M_{D_s}^2-M_{\phi}^2-q^2)(M_{D_s}+M_{\phi})A_1(q^2) \nonumber \\
&-& 4 \frac{M_{D_s}^2p_{\phi}^2}{M_{D_s}+M_{\phi}}A_2(q^2) ] . \nonumber
\end{eqnarray}
The $p_{\phi}q^2$ factor in the differential cross-section (Eq.~\ref{eq:diffrate}) 
means that 
the contribution of $H_0(q^2)$ does not diverge as $q^2 \rightarrow 0$. 
Note that $A_1$ contributes to all helicities, $A_2$ only to $H_0$ and 
$V$ only to $H_{\pm}$. At high $q^2$ where $p_{\phi} \rightarrow 0$, $A_1$ 
dominates all of the helicities. 

In the differential decay rate given in Eq.~\ref{eq:diffrate}, the lepton mass is neglected.
This is a good approximation for $D_s \to \phi$ semileptonic decays 
where $\ell = e, \mu$ and one which we make for our comparison 
with BaBar results~\cite{babar}, in which the final state lepton is an electron.

However, we can calculate in lattice QCD the contributions to 
the decay rate that 
are suppressed by factors of $m_\ell^2$ and study their 
relative size.
If we do not neglect the lepton mass, the decay rate also includes~\cite{Korner:1989qb}
\begin{align}
&\frac{3}{8(4\pi)^4}G_F^2|V_{cs}|^2\frac{p_{\phi}m_\ell^2}{M^2_{D_s}}{\mathcal{B}}(\phi \rightarrow K^+K^-) \times \label{eq:mlsqdiffrate} \\
 \left\{ \right.& \sin^2\theta_K \sin^2\theta_\ell |H_+(q^2)|^2 \nonumber \\
 +& \sin^2\theta_K \sin^2\theta_\ell |H_-(q^2)|^2 \nonumber \\
 +& 4\cos^2\theta_K\cos^2\theta_\ell |H_0(q^2)|^2 \nonumber \\
 +& 4\cos^2\theta_K |H_t(q^2)|^2 \nonumber \\
 +& \sin^2\theta_K\sin^2\theta_\ell \cos2\chi H_+(q^2)H_-(q^2) \nonumber \\
 +& \sin2\theta_K \sin2\theta_\ell \cos2\chi H_+(q^2)H_0(q^2) \nonumber \\
 +& \sin2\theta_K \sin2\theta_\ell \cos2\chi H_-(q^2)H_0(q^2) \nonumber \\
 +& 2\sin2\theta_K\sin\theta_\ell\cos\chi H_+(q^2)H_t(q^2) \nonumber \\
 +& 2\sin2\theta_K\sin\theta_\ell\cos\chi H_-(q^2)H_t(q^2) \nonumber \\
 +& \left. 8\cos^2\theta_K\cos\theta_\ell H_0(q^2)H_t(q^2) \right\}. \nonumber 
\end{align}
All of the cross terms in Eq.~\ref{eq:mlsqdiffrate} vanish on integration over $\chi$, apart from $H_0(q^2)H_t(q^2)$, which vanishes if we integrate over $\cos\theta_\ell$.

The helicity amplitude $H_t(q^2)$ is given by 
\begin{equation} H_t(q^2) = \frac{2 M_{D_s}p_\phi}{\sqrt{q^2}}A_0(q^2). \end{equation}
At $q^2=0$, $H_t(0) = H_0(0)$ because, for these kinematics, $M_{D_s}^2 - M_\phi^2 = 2 M_{D_s} p_\phi$ and we also have $A_0(0) = A_3(0)$.
We can calculate $A_0(q^2)$ using a pseudoscalar current (see Eq.~\ref{eq:pme}), so it 
is straightforward to calculate $H_t(q^2)$ in lattice QCD.

As $H_t$ is proportional to $1/\sqrt{q^{2}}$, it is most important at low $q^2$.
The effect of this helicity amplitude could be detected as a difference in the semileptonic decay rate with electrons or muons in the final state.
It has been observed in the measurements of 
$D \to K^* \ell \nu$ made by CLEO \cite{cleodkstar}. 

\section{Lattice Calculation.} 
\label{sec:lattice}

For the lattice QCD calculation we use the Highly Improved Staggered 
Quark action~\cite{HISQ_PRD} for all the valence quarks. This action 
has very small discretisation errors, making it an excellent action 
for $c$~\cite{HISQ_PRD, HISQ_PRL, Dsdecayconst, jpsi}  
as well as for the lighter $s$ quarks we need here. 
We calculate HISQ propagators on gluon field configurations generated 
by the MILC collaboration that include $u$, $d$ and $s$ sea quarks 
using the asqtad formalism~\cite{MILCconfigs}. Table~\ref{tab:params} 
gives the parameters of the ensembles of configurations we use, with two different 
lattice spacing values and two different $u/d$ sea quark masses.  

\begin{table*}
\begin{tabular}{llllllllll}
\hline
\hline
Set &  $r_1/a$ & $au_0m_{l}^{asq}$ & $au_0m_{s}^{asq}$ & $m_l/m_{s,phys}$ & $L_s/a \times L_t/a$ & $am_{s}^{HISQ}$ & $am_{c}^{HISQ}$ & $n_{cfg}$ & $T$ \\
\hline
1 &  2.647(3) & 0.005 & 0.05 & 0.14 & 24 $\times$ 64 & 0.0489 & 0.622 & 2088 & 12, 15, 18 \\
2 &  2.618(3) & 0.01 & 0.05 & 0.29 & 20 $\times$ 64 & 0.0496 & 0.63 & 2259 & 12, 15, 18 \\
\hline
3 & 3.699(3) & 0.0062 & 0.031 & 0.24 & 28 $\times$ 96 & 0.0337 & 0.413 & 1911 & 16, 19, 20, 23\\
\hline 
\hline
\end{tabular}
\caption{Ensembles (sets) of MILC configurations used here. 
Sea 
(asqtad) quark masses $m_{\ell}^{asq}$ ($\ell = u/d$) and $m_s^{asq}$ 
use the MILC convention where $u_0$ is the plaquette 
tadpole parameter. 
The lattice spacing is given in units of $r_1$ after `smoothing'
~\cite{MILCconfigs}. We use $r_1=0.3133(23)$ fm~\cite{Davies:2009tsa}. 
Sets 1 and 2 are `coarse' ($a \approx 0.12$ fm) and set 3, 
`fine' ($a \approx 0.09$ fm).  The lattice size 
is given by $L_s^3 \times L_t$. 
Column 5 gives the sea light quark mass in units of 
the physical strange mass, as determined in~\cite{Dsdecayconst}. 
Columns 7 and 8 give the valence $s$ and $c$ HISQ quark masses, 
tuned to the 
physical values~\cite{Dsdecayconst}. 
We use 4 time sources on each of the $n_{cfg}$ configurations. 
The final column lists the $T$ values used in the 3-pt 
correlators (see Fig.~\ref{fig:3ptdiag}). 
}
\label{tab:params}
\end{table*}

To tune the $s$ and $c$ quark masses to 
their correct physical values we use the pseudoscalar $\eta_s$ 
and $\eta_c$ meson masses~\cite{Dsdecayconst}.
The $\eta_s$ is a fictitious $s\overline{s}$ pseudoscalar 
that is not allowed to decay in lattice QCD.
Although this meson does not occur in the real 
world its mass can be accurately determined in 
lattice QCD because it does not contain valence 
$u/d$ quarks, and a `physical' value for its mass can be
determined in the continuum and chiral limits. 
We find $M_{\eta_s}$ = 0.6858(40) GeV~\cite{Davies:2009tsa}, 
and use this to tune the $s$ quark mass~\cite{Dsdecayconst}. 
In tuning the $c$ quark mass here we must use the 
value of the $\eta_c$ mass~\cite{Dsdecayconst} in a world without 
electromagnetism or $c$ quarks in the sea. We 
take this to be $M_{\eta_c}$=2.985(3) GeV~\cite{Gregory:2010gm}. 
Discretisation errors from using the HISQ action 
are reduced for $c$ quarks by modifying the coefficient 
of the `Naik' term~\cite{naik}, which corrects for 
$a^2$ errors in the covariant derivative, to 
include the tree-level correction 
which is a function of the bare quark 
mass, $m_ca$~\cite{Dsdecayconst}. 
A measure of the smallness of the resulting discretisation 
errors comes from a study of the `speed of light' 
for the $\eta_c$~\cite{jpsi}. This differs from 1 by 
less than 3\% on both the coarse and fine lattices. 

\begin{figure}
\centering
\includegraphics[width=0.45\textwidth]{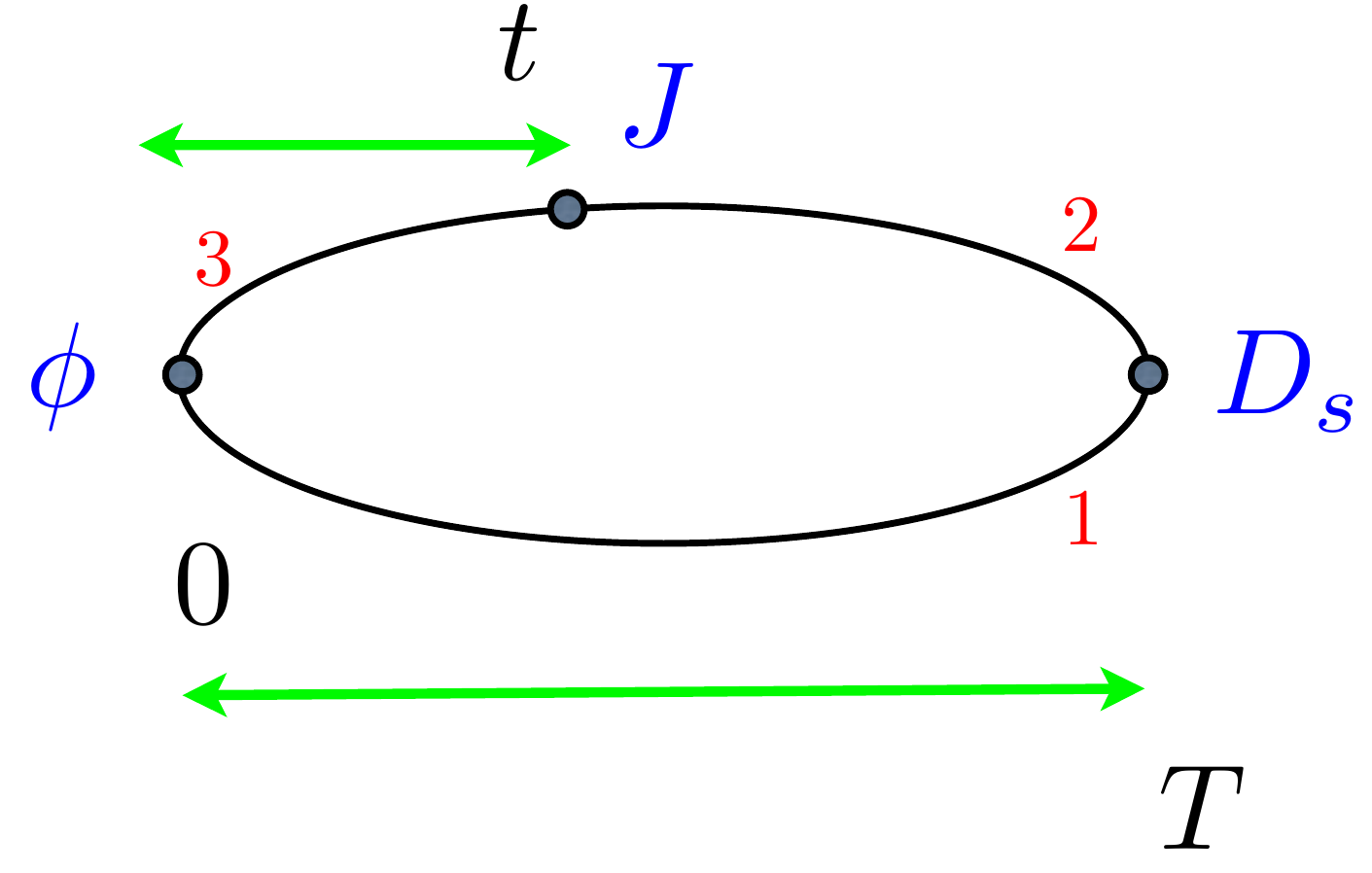}
\caption{A schematic diagram of the 3-point function for $D_s \rightarrow \phi \ell \nu$ decay. Different currents, $J$, are inserted at the vertex, as 
described in the text, to obtain specific form factors.}
\label{fig:3ptdiag}
\end{figure}

The quark propagators are made from a `random wall' 
source - a colour-vector of random numbers in U(1) on a 
source timeslice - to reduce the statistical noise. 
We use four
evenly spaced time sources on each configuration, choosing 
the first time source randomly to reduce correlations 
between configurations. 

The lattice spacing is determined for each ensemble using 
the calculation of static quark potential by MILC and the 
extraction of a parameter associated with that potential 
called $r_1$~\cite{MILCconfigs}. The value for this 
parameter in units of the lattice spacing, $r_1/a$, is 
given in Table~\ref{tab:params}. Using a physical value 
for $r_1$, $r_1$ = 0.3133(23) fm~\cite{Davies:2009tsa}, 
allows us to convert these numbers to a value for 
the lattice spacing, $a$, on each ensemble. This allows 
us in turn to convert all dimensionful quantities calculated
on the lattice into GeV units. 

The HISQ $s$ and $c$ 
quark propagators calculated on these gluon field configurations 
are combined to 
make meson correlators (2-point functions) for 
$D_s$ and $\phi$ and 3-point functions that allow 
us to calculate the $D_s$ to $\phi$ transition matrix 
element. 

Correlators for mesons with specific spin-parity quantum 
numbers are made using staggered quarks (which have no 
spin degree of freedom) by including space-time-dependent 
phases of $\pm 1$ at source and sink. This can be combined 
with a point-splitting of the source/sink operator.  
Because of fermion doubling there are 16 `tastes' of every 
meson. 
We will use the spin-taste notation $\gamma_n \otimes \gamma_s$ to denote a staggered bilinear with spin $\gamma_n$ and taste $\gamma_s$.
The masses of the different tastes differ by 
discretisation errors (at $\mathcal{O}(a^2)$) and 
we are free to use whichever taste is the most convenient 
for each calculation. We will make use of that freedom here. 
However, because point-split source and sink operators typically give noisier 
results than local operators, we will restrict ourselves 
to at most a 1-link point-splitting. 

We will use two different local operators to create/destroy 
$D_s$ mesons. One is the local $\gamma_5$ `Goldstone' operator 
(i.e. $\gamma_5 \otimes \gamma_5$) 
and the other is the local $\gamma_t\gamma_5$ operator.  
The pattern of taste-splittings for pseudoscalar mesons is 
well mapped out and significantly reduced for HISQ quarks~\cite{HISQ_PRD} 
over those in the asqtad formalism~\cite{milc-hisq}. 
We expect splittings in the squared-mass of the different 
pseudoscalar tastes to be proportional to $a^2$. 
The lightest pseudoscalar is the Goldstone meson and 
the next lightest is that of the local $\gamma_t\gamma_5$ 
operator. Since the $a^2$ taste-splitting effect is in the square of the mass, 
the mass splitting between pseudoscalars actually falls 
as the mass increases (as long as the mass does not become 
too large)~\cite{HISQ_PRD}. So in fact the difference in mass between 
the two tastes of $D_s$ used here is very small, as 
we will see in Section~\ref{sec:results}. 

For vector mesons taste-splittings are significantly 
smaller~\cite{jpsi}. Here we will use both a local and 
a 1-link point-split operator for the $\phi$ and 
discuss results from those and the comparison between 
them in~\ref{subsec:phi}.  
In principle the $\phi$ meson is a flavor-singlet. However, 
we expect the effect of `disconnected' diagrams (two $s$ quark 
loops connected only by gluon exchange) to be small for 
vector mesons~\cite{McNeile:2001cr} and we do not include them here. 
In the real world the $\phi$ decays strongly to $K\overline{K}$ 
but not in our lattice QCD simulations. We consider the 
effect of that on our $\phi$ mesons in Section~\ref{subsec:phi}.  

A schematic diagram for the 3-point function for $D_s$ to $\phi$ 
decay is shown in Fig.~\ref{fig:3ptdiag}.  
Quark propagators 1 and 3 correspond to $s$ quarks and 
propagator 2 is for a $c$ quark. Propagators 1 and 3 
are tied together with appropriate phases to make 
a $\phi$ meson at the origin. Propagator 2 is 
calculated from a source 
made from propagator 1 at timeslice $T$, using appropriate 
phases for a pseudoscalar $D_s$ meson.  
Finally propagators 2 and 3 are combined at timeslice 
$t$ with appropriate phases to correspond to a vector, 
axial vector or pseudoscalar current, so that we 
can determine the vector and axial vector form factors 
discussed in Section~\ref{sec:theory}. 

To cover the range of squared 4-momentum transfer, $q^2$, 
available in the decay 
we keep the $D_s$ meson at rest and give spatial 
momentum to the $\phi$ meson varying from zero up to 
an appropriate value to set $q^2=0$. 
We do this by calculating $s$ quark propagators 
for propagator 3 that carry spatial momentum through 
the use of a `twisted boundary condition'~\cite{firsttwist, etmctwist}.
If propagator 3 is calculated with boundary condition 
\begin{equation}
\chi(x + \hat{e}_j L) = e^{ i \theta_j} \chi(x),
\label{eq:twist}
\end{equation}
then the momentum of the $\phi$ meson made 
by combining propagators 1 and 3 with our random wall 
sources and summing over spatial sites at the sink is
\begin{equation}
p_j = \frac{\theta_j}{L_s}.  
\label{eq:momtwist}
\end{equation}
The boundary condition in eq.~(\ref{eq:twist}) is 
actually implemented by multiplying the gluon links 
in the $j$ direction by phase $\exp(i \theta_j/L_s)$. 

The 3-point function for $D_s \rightarrow \phi$ is 
calculated for all $t$ values from 
$0$ to $T$ and for several values of $T$ (which include both even and odd values as given 
in Table~\ref{tab:params}) so that the dependence 
of the function on $t$ and $T$ can be fully mapped 
out. The 3-point function is fit simultaneously 
with the 2-point function using fit forms  
\begin{eqnarray}
C^{(P)}_{2pt} &=& \sum_{i_n,i_o} \{d^{(P)}_{i_n}\}^2 \mathrm{fn}(E^{(P)}_{i_n},t^{\prime}) - \{{\tilde{d}}^{(P)}_{i_o}\}^2 \mathrm{fo}(\tilde{E}^{(P)}_{i_o},t^{\prime}) \nonumber \\
C^{P\rightarrow Q}_{3pt} &=& \sum_{i_n,j_n} d^{(P)}_{i_n} \mathrm{fn}(E^{(P)}_{i_n},t) J^{nn}_{i_n,j_n} d^{(Q)}_{j_n} \mathrm{fn}(E^{(Q)}_{j_n},T-t) \nonumber \\
&-&\sum_{i_n,j_o} d^{(P)}_{i_n} \mathrm{fn}(E^{(P)}_{i_n},t) J^{no}_{i_n,j_o} \tilde{d}^{(Q)}_{j_o} \mathrm{fo}({\tilde{E}}^{(Q)}_{j_o},T-t) \nonumber \\
&+&( n \leftrightarrow o)  
\label{eq:3ptfit}
\end{eqnarray}
with 
\begin{eqnarray}
\mathrm{fn}(E,t) &=& e^{-Et} + e^{-E(L_t-t)} \nonumber \\
\mathrm{fo}(E,t) &=& (-1)^{t/a} \mathrm{fn}(E,t) . 
\label{eq:fnfo}
\end{eqnarray}
We use Bayesian methods~\cite{gplbayes} 
that allow us to include the effect of excited states, both 
`radial' excitations ($n$) and, because we are using staggered quarks, 
opposite parity mesons that give oscillating terms ($o$). 
We fit all the 2-point and 3-point correlators on a given ensemble 
at multiple momenta 
simultaneously to take account of correlations. 
The Bayesian approach requires the constraint of prior values and 
widths on the parameters. These are taken as: 
ground-state energy, 2\% width; splitting 
between ground-state and excited energies, 600 MeV with 50\% width; splitting 
between ground-state and lowest oscillating state, 400 MeV with 50\% width; amplitudes, 0.01(1.0) for normal states 
and 0.01(0.5) for oscillating states; matrix elements, 0.01(1.0).

In Eq.~\ref{eq:3ptfit}, $d_{i_n}$ are the 
amplitudes for creation/annihilation of 
the $D_s$ or $\phi$ mesons. The amplitude can be converted 
into the decay constant and this will be discussed 
for the $\phi$ in Section~\ref{subsec:phi}. 
Results for the $D_s$ mass and decay constant on these gauge configurations were presented in \cite{Dsdecayconst}.
$J_{i_n,j_n}$ is related to the matrix element of 
the vector, axial vector or pseudoscalar current 
between $D_s$ and $\phi$. By matching to a continuum 
correlator with a relativistic normalisation of states and allowing 
for a renormalisation of the lattice current we see that the 
matrix elements between the ground state mesons that we want to determine
are given by 
\begin{equation}
\langle D_s |J| \phi \rangle = Z\sqrt{4E^{(D_s)}_{0}E^{(\phi)}_{0}}J^{nn}_{0,0}.
\label{eq:me}
\end{equation}

The vector current we use for the $D_s \to \phi$ transition is a local spatial current.
We use both a local and a point-split axial vector current. 
The point-split current does not include gauge links because we work in the Coulomb gauge.
The local pseudoscalar current we use is absolutely normalised when multiplied by the lattice quark mass.
The vector and axial vector currents are nonperturbatively normalised, as described in Appendix \ref{sec:zfactors}, and the $Z$ factors we obtain on each of our ensembles are given in Table \ref{tab:csz}.

\section{Results} 
\label{sec:results}

\subsection{The $\phi$ meson}
\label{subsec:phi}

\begin{table*} 
\begin{center} 
\begin{tabular}{cccccccc} 
\hline 
\hline 
Set & $am_s$ & $aM_{\eta_s}$ & $aM_{\phi} (\gamma_\mu \otimes \gamma_{\mu})$ & $af_{\phi}/Z (\gamma_\mu \otimes \gamma_{\mu})$ & $aM_{\phi} (\gamma_\mu \otimes 1)$ & $af_{\phi}/Z (\gamma_\mu \otimes 1)$ \\ 
\hline 
1 & 0.0489 & 0.4111(1) & 0.6386(26) & 0.1504(36) & 0.6365(44) & 0.1341(59) \\ 
\hline 
2 & 0.0496 & 0.4163(1) & 0.6549(31) & 0.1571(41) & 0.6569(28) & 0.1401(30) \\ 
\hline 
3 & 0.0337 & 0.2937(1) & 0.4550(35) & 0.1100(30) & 0.4570(21) & 0.1026(34) \\ 
\hline 
\hline 
\end{tabular} 
\caption{For each ensemble we give the $s$ valence quark masses 
and $\eta_s$ meson mass in lattice units. These are followed by: 
(columns 4 and 5) the mass and bare (unrenormalised) decay constant for the local $\phi$ and 
(columns 6 and 7) the mass and bare decay constant for the 1-link $\phi$. 
}
\label{tab:phi} 
\end{center} 
\end{table*}

When handling the $\phi$ meson in our lattice QCD calculations 
we have treated it as a pure $s\overline{s}$ vector meson and 
not included quark-line disconnected diagrams that could mix 
in light-quark components. These effects are expected to 
be very small from phenomenology. For example the width for 
$\phi$ to decay to $\pi^0\gamma$, which would be zero for 
a pure $s\overline{s}$ $\phi$, is 5.5keV~\cite{pdg} (branching fraction 0.13\%). 
This compares to a width to $\pi^0\gamma$ for the light vector 
$\omega$ of 700keV~\cite{pdg} (branching fraction 8\%).  
In lattice QCD calculations where quark-line disconnected diagrams 
have been included they are indeed found to have tiny effect 
for vectors. Ref.~\cite{Dudek:2011tt} gives a mixing angle between 
$\phi$ and $\omega$ of $1.7(2)^{\circ}$ for relatively heavy 
light quarks. We conclude that quark-line disconnected diagrams 
are a negligible issue here. 

The $\phi$ meson in the real world 
decays strongly to $K\overline{K}$ and hence is not `gold-plated'. 
The $\phi$ meson mass is close to threshold for this decay, 
however, and so the $\phi$ width is small (4 MeV~\cite{pdg}).   
It may then be true that the impact of the decay channel is 
not large and it may effectively be possible to treat the
$\phi$ as being close to gold-plated within lattice QCD. 

A simple model by which we can analyse the effect of the $K\overline{K}$ 
channel on the $\phi$ is to treat both $\phi$ and $K$ as elementary 
particles and couple them with a P-wave vertex, $g\varepsilon \cdot p$. 
Here $\varepsilon$ is the polarization vector of the $\phi$ and 
$p$ is the momentum of the $K$ in the $\phi$ rest frame. Then, 
from perturbation theory treating the $K$ as nonrelativistic
\begin{equation}
\Delta E_{\phi} = g^2 \int^{\Lambda} \frac{d^3p}{(2\pi)^3} \frac{|\varepsilon \cdot p|^2}{\Delta M - p^2/M_K + i\epsilon}
\label{eq:mshift}
\end{equation}
where $\Delta M \equiv M_{\phi} - 2M_K$.
Spin-averaging, and absorbing factors into the coupling constant, 
gives  
\begin{eqnarray}
\Delta E_{\phi} &=& \tilde{g}^2 \int^{\Lambda}_0 dp \frac{p^4}{\gamma^2 - p^2 + i\epsilon} \nonumber \\
&=& \tilde{g}^2 (-\frac{\Lambda^3}{3} - \gamma^2\Lambda - \frac{i\pi}{2}\gamma^3 )
\label{eq:mshift2}
\end{eqnarray}
dropping higher order terms in $\gamma$ where $\gamma^2 = (M_{\phi}-2M_K)M_K$.
The imaginary part is the width, $\Gamma_{\phi}=\tilde{g}^2\pi\gamma^3$, and we can 
use this to estimate $\tilde{g}$. Using the physical width and physical masses~\cite{pdg} 
gives $\tilde{g}^2 \approx 0.35$ for both charged $K$ and neutral $K$ decay modes. 
The shift in the mass expected from coupling to the $K$ decay mode is 
then $-\tilde{g}^2\gamma^2\Lambda$, giving a result of $\approx - 5 \mathrm{MeV}$.
This is a very small effect, less than 0.5\% of the mass, so that 
the $\phi$ meson behaves as a gold-plated particle to a good 
approximation. 

On the lattice, the coupling to the $K\overline{K}$ decay mode is 
distorted by the fact that the $K$ meson will typically have a higher
mass than its physical value because the sea $u/d$ quarks will be too
heavy. Then $M_{\phi}-2M_K$ will be negative and the expected shift 
in the $\phi$ meson mass resulting from coupling to $K\overline{K}$ will 
be positive. Thus we expect the $\phi$ meson mass on the lattice to 
have more dependence on the sea $u/d$ quark masses than a typical 
gold-plated meson would have, and for its mass to be too high for 
unphysically heavy $u/d$ quark masses.  

\begin{figure}
\centering
\includegraphics[width=0.45\textwidth]{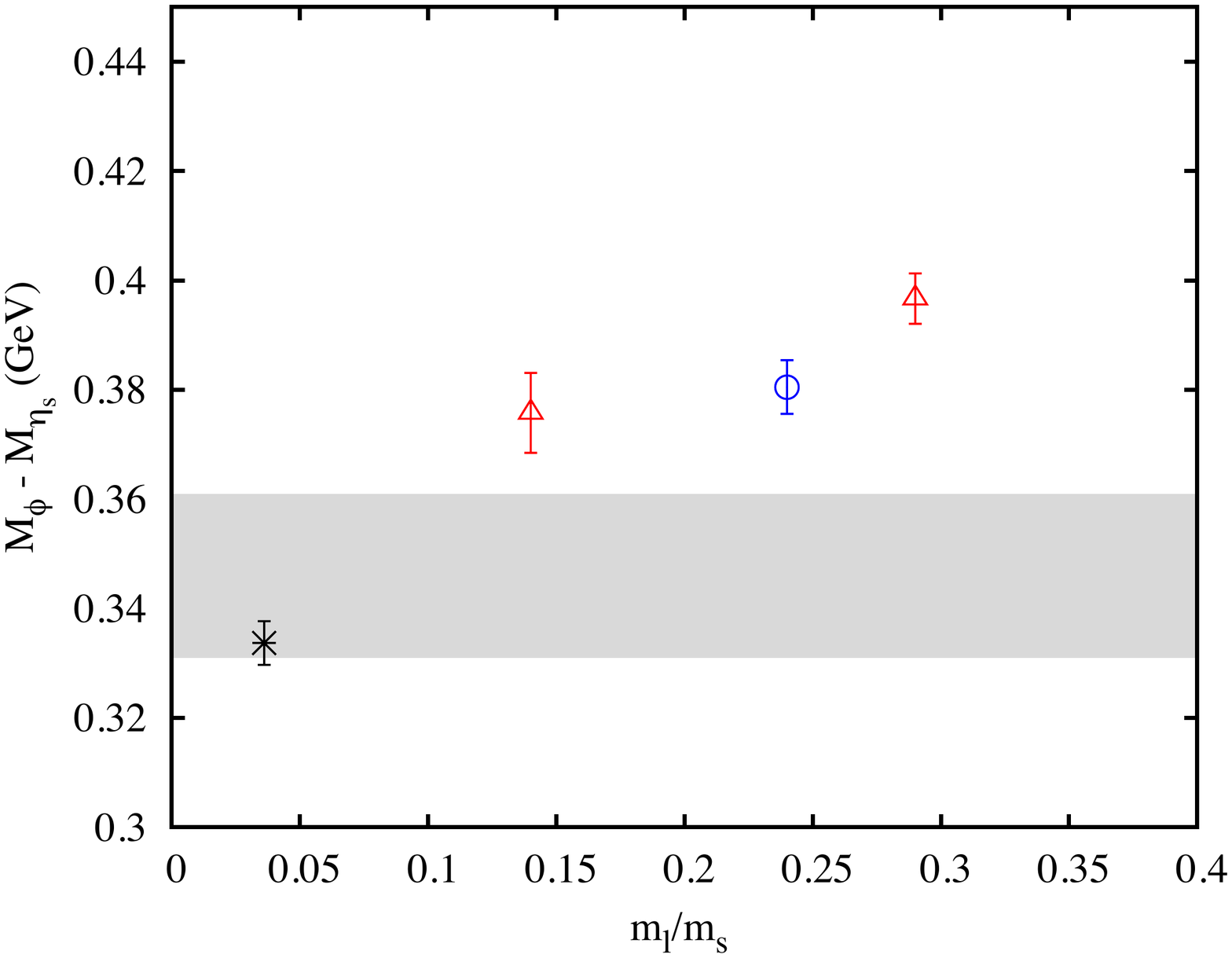}
\includegraphics[width=0.45\textwidth]{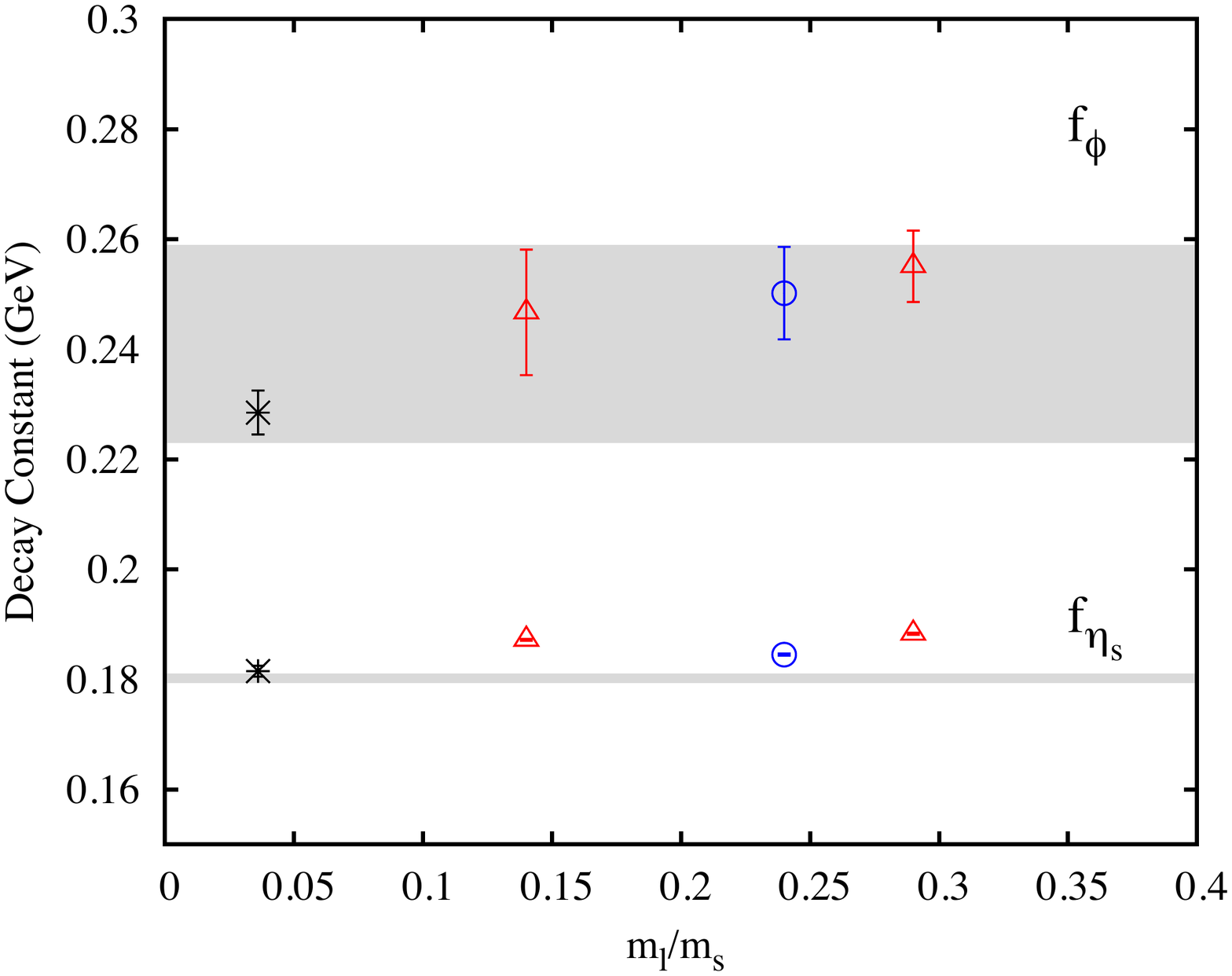}
\caption{
Upper plot: The difference in mass between the $s\overline{s}$ vector $\phi$ and pseudoscalar 
$\eta_s$ as a function of sea light quark mass in units of the 
physical strange quark mass for coarse lattices (red open triangles) and 
fine (green open circle). The `experimental' result is plotted with a 
black burst. This is obtained from the experimental result for $M_{\phi}$ 
and the lattice QCD result for $M_{\eta_s}$ in the continuum and 
chiral limits~\cite{Davies:2009tsa}. The gray shaded band gives the lattice 
result in the continuum and chiral limits from a simple fit described 
in the text.  \\
Lower plot: A similar plot for the $\phi$ and $\eta_s$ 
decay constants. The black bursts denote the experimental
result for the $\phi$ obtained from its leptonic width~\cite{pdg} 
and the result from lattice QCD for the $\eta_s$~\cite{Davies:2009tsa}.
}
\label{fig:mphi}
\end{figure}

Both of these effects are borne out in our results. 
Table~\ref{tab:phi} 
gives results for the $\phi$ mass obtained from the combined 
fit to 2-point and 3-point functions described in Eq.~\ref{eq:3ptfit}, 
along with values of the $\eta_s$ meson mass for the 
same valence $s$ quarks. 
Notice that the statistical error on the $\phi$ mass is much 
larger than that for the $\eta_s$. This is a consequence of 
the exponentially falling signal-to-noise ratio for particles 
like the $\phi$ where the noise amplitude is governed by a lighter 
mass (in this case the $\eta_s$) than the signal. 
Figure~\ref{fig:mphi} 
shows the difference in mass between the $\phi$ and the $\eta_s$ 
plotted as a function of the sea light quark mass in units of the 
physical strange quark mass (which is given in Table~\ref{tab:params}). 
The figure shows results for the 1-link $\phi$ operator but results 
for the local $\phi$ are similar. 
The shaded band gives the result at the physical value of $m_l/m_s$ 
(1/27.5 ~\cite{pdg}) from a simple fit. 
This allows for a linear term in $m_l/m_s$ and quadratic and 
quartic terms in $a$. We obtain the result 0.346(15) GeV at the physical 
point, 
which agrees within errors with the value obtained from the 
experimental result for the $\phi$ mass and the lattice result 
for the $\eta_s$ mass in the continuum and 
chiral limits~\cite{Davies:2009tsa} (0.3337(40) GeV). 
Hence it seems that light quark mass effects 
from the coupling to $K\overline{K}$ are relatively benign 
compared to the significant statistical errors that 
we have in the $\phi$ mass. 

A similar result is seen for the $\phi$ meson decay constant, $f_{\phi}$. 
Table~\ref{tab:phi} 
gives results for the $\phi$ decay constant obtained from the combined 
fit to 2-point and 3-point functions described in Eq.~\ref{eq:3ptfit}. 
The decay constant is extracted from the amplitude of the ground-state 
$\phi$ in the 2-point 
function, $d^{\phi}_0$, using 
\begin{equation}
f_{\phi} = Z d^{\phi}_0\sqrt{\frac{2}{E^{\phi}_0}}
\label{eq:fphi}
\end{equation}
where $Z$ is the renormalisation factor required to 
match the lattice vector current used to create or destroy 
the $\phi$ to the continuum. We give results for both the 
local vector current and the 1-link vector current in Table \ref{tab:phi} and they 
can both be renormalised fully nonperturbatively. How this is 
done is described in Appendix~\ref{sec:zfactors} and the appropriate renormalisation constants 
are given in Table~\ref{tab:csz}.

We plot $f_\phi$ in Figure~\ref{fig:mphi} for the 1-link $\phi$.
Results for the local $\phi$ are similar. 
A simple extrapolation, as for the mass, 
gives a physical result of $f_\phi = 241(18)$ MeV, which agrees 
within errors with the
experimental value of $f_\phi = 229(4)$ MeV obtained 
from $\Gamma(\phi\to e^+e^-)$ (1.27(4) keV~\cite{pdg}). 
The decay constant is related to the leptonic decay rate by
\begin{equation} \Gamma(\phi\to e^+e^-) = \frac{4\pi}{3}\alpha^2_{QED} e_s^2\frac{f_\phi^2}{m_\phi}, \end{equation}
where $e_s$ is the electric charge of the $s$ quarks in 
units of $e$, i.e. $-1/3$.  
We show for comparison the results for the decay constant of 
the $\eta_s$ which was studied on the lattice in~\cite{Davies:2009tsa}. 
There is some sign that the $\phi$ meson has enhanced 
dependence on $m_l/m_s$ compared to that of the 
gold-plated $\eta_s$. This might be expected to be a result of coupling 
to $K\overline{K}$ but it causes no problem in the 
extrapolation to the physical point of the decay 
constant within the lattice errors of 7\%. 

For a phenomenological comparison that could shed light 
on how large an effect we might expect on meson 
properties from coupling to decay channels, we
can compare the $\omega$ and the $\rho$. They are both made 
of light $u/d$ valence quarks but the $\rho$ has a 
strong two-body decay mode to $\pi\pi$ which is not allowed 
for the $\omega$ by G-parity (which instead decays to three
$\pi$). The $\omega$ and $\rho$ masses nevertheless agree 
to within 10 MeV so little effect of the $\rho$ decay mode is seen there. 
We expect the leptonic width of the $\omega$ to be one-ninth 
that of the $\rho$ simply from isospin~\cite{close}. 
In fact this expectation is violated by about 30\%, which might 
indicate a 15\% effect in the $\rho$ decay constant from coupling 
to decay channels. This should be compared to a 7(7)\% possible effect in 
the $\phi$ from our results, as described above.  
  
For possible effects on a meson to meson transition 
rate where one of the mesons in the process is 
gold-plated and the other 
is not, we can compare the 
decay of $\rho$ and $\omega$ to $\pi^0\gamma$. Here we 
expect the rate for the $\rho$ to be one-ninth that of the 
$\omega$~\cite{close}. This expectation is violated 
by 12\%, indicating a possible 6\% effect in the matrix element.  
Following the discussion of the decay constant above, 
this would mean that a reasonable error to take 
on a transition matrix element involving the $\phi$ (i.e. 
$D_s \rightarrow \phi \ell \nu$) as a result of 
$\phi$ coupling to $K\overline{K}$ might be half this, i.e. 3\%. 

\subsection{Results for form factors} 
\label{subsec:resff}

As discussed in Section~\ref{sec:lattice} we calculate 3-point functions 
for $D_s$ to $\phi$ decay by inserting either a vector, axial vector 
or pseudoscalar current between the $D_s$ and the $\phi$. By choosing 
appropriate kinematic conditions we can isolate the individual 
form factors from Eq.~\ref{eq:ffdef}. We keep the $D_s$ meson at 
rest but can give the $\phi$ spatial momentum in different directions 
and choose its spin polarization. 

For a 3-point function made of staggered quarks all the tastes must `cancel'. 
Therefore only certain tastes of current can be used with certain tastes 
of mesons. Below we discuss each of the form factors we extract, explaining 
the method used. The $Z$ factors for all of the different currents are determined 
fully non-perturbatively and we describe how that is done in Appendix~\ref{sec:zfactors}. 
Table~\ref{tab:dsphidata} collects all of the form factor results. 

\begin{table*}
\begin{tabular}{cccccccc}
\hline
\hline
Set & Form factor & T values & $am_{D_s}$ & $\theta$ & $aE_\phi$ & $a^2q^2$ & $F(q^2)$ \\
\hline
1 & Local $A_1(q^2)$ & 12,15,18 & 1.1889(1) & 0.0 & 0.6386(26) & 0.303(3) & 0.685(8) \\
1 & Local $A_1(q^2)$ & 12,15,18 & 1.1889(1) & 7.0 & 0.7108(40) & 0.143(4) & 0.657(12) \\
1 & Local $A_1(q^2)$ & 12,15,18 & 1.1889(1) & 10.18 & 0.7817(62) & -0.014(5) & 0.624(22) \\
1 & 1-link $A_1(q^2)$ & 12,15,18 & 1.1889(1) & 0.0 & 0.6365(44) & 0.305(6) & 0.694(17) \\
1 & 1-link $A_1(q^2)$ & 12,15,18 & 1.1889(1) & 7.0 & 0.7056(51) & 0.148(5) & 0.648(25) \\
1 & 1-link $A_1(q^2)$ & 12,15,18 & 1.1889(1) & 10.18 & 0.7815(82) & -0.014(6) & 0.642(34) \\
1 & $A_0(q^2)$ & 12,15,18 & 1.1889(1) & 7.0 & 0.7090(35) & 0.145(3) & 0.808(24) \\
1 & $A_0(q^2)$ & 12,15,18 & 1.1889(1) & 10.18 & 0.7784(51) & -0.011(4) & 0.707(23) \\
1 & $A_2(q^2)$ & 12,15,18 & 1.1889(1) & 7.0 & 0.7090(35) & 0.145(3) & 0.529(106) \\
1 & $A_2(q^2)$ & 12,15,18 & 1.1889(1) & 10.18 & 0.7784(51) & -0.011(4) & 0.430(90) \\
1 & $V(q^2)$ & 12,15,18 & 1.1909(5) & 7.0 & 0.7121(41) & 0.144(3) & 1.141(72) \\
1 & $V(q^2)$ & 12,15,18 &  1.1909(5) & 10.18 & 0.7853(61) & -0.015(5) & 1.055(69) \\
\hline
2 & Local $A_1(q^2)$ & 12,15,18 & 1.2015(1) & 0.0 & 0.6549(31) & 0.298(3) & 0.689(13) \\
2 & Local $A_1(q^2)$ & 12,15,18 & 1.2015(1) & 6.0 & 0.7195(56) & 0.142(5) & 0.627(26) \\
2 & Local $A_1(q^2)$ & 12,15,18 & 1.2015(1) & 8.39 & 0.7769(72) & 0.004(6) & 0.596(28) \\
2 & 1-link $A_1(q^2)$ & 12,15,18 & 1.2015(1) & 0.0 & 0.6569(28) & 0.296(3) & 0.684(11) \\
2 & 1-link $A_1(q^2)$ & 12,15,18 & 1.2015(1) & 6.0 & 0.7246(54) & 0.137(5) & 0.627(23) \\
2 & 1-link $A_1(q^2)$ & 12,15,18 & 1.2015(1) & 8.39 & 0.7713(100) & 0.009(9) & 0.582(42) \\
2 & $A_0(q^2)$ & 12,15,18 & 1.2015(1) & 6.0 & 0.7151(49) & 0.146(5) & 0.787(29) \\
2 & $A_0(q^2)$ & 12,15,18 & 1.2015(1) & 8.39 & 0.7825(48) & -0.001(4) & 0.701(22) \\
2 & $A_2(q^2)$ & 12,15,18 & 1.2015(1) & 6.0 & 0.7151(49) & 0.146(5) & 0.475(183) \\
2 & $A_2(q^2)$ & 12,15,18 & 1.2015(1) & 8.39 & 0.7825(48) & -0.001(4) & 0.345(107) \\
2 & $V(q^2)$ & 12,15,18 & 1.2040(3) & 6.0 & 0.7176(67) & 0.146(6) & 1.116(87) \\
2 & $V(q^2)$ & 12,15,18 & 1.2040(3) & 8.39 & 0.7738(82) & 0.009(7) & 1.057(180) \\
\hline
3 & Local $A_1(q^2)$ & 16,19,20,23 & 0.8460(1) & 0.0 & 0.4550(35) & 0.153(3) & 0.717(23) \\
3 & Local $A_1(q^2)$ & 16,19,20,23 & 0.8460(1) & 6.0 & 0.5007(34) & 0.073(2) & 0.635(12) \\
3 & Local $A_1(q^2)$ & 16,19,20,23 & 0.8460(1) & 8.39 & 0.5563(54) & 0.000(3) & 0.648(23) \\
3 & 1-link $A_1(q^2)$ & 16,19,20,23 & 0.8460(1) & 0.0 &  0.4570(21) & 0.151(2) & 0.716(11) \\
3 & 1-link $A_1(q^2)$ & 16,19,20,23 & 0.8460(1) & 6.0 & 0.5038(33) & 0.071(2) & 0.658(14) \\
3 & 1-link $A_1(q^2)$ & 16,19,20,23 & 0.8460(1) & 8.39 & 0.5463(43) & 0.000(6) & 0.638(18) \\
3 & $A_0(q^2)$ & 16,19,20,23 & 0.8460(1) & 6.0 & 0.4960(54) & 0.076(4) & 0.783(51) \\
3 & $A_0(q^2)$ & 16,19,20,23 & 0.8460(1) & 8.39 & 0.5523(20) & -0.003(1) & 0.689(12) \\
3 & $A_2(q^2)$ & 16,19,20,23 & 0.8460(1) & 6.0 & 0.4960(54) & 0.076(4) & 0.499(181) \\
3 & $A_2(q^2)$ & 16,19,20,23 & 0.8460(1) & 8.39 & 0.5523(20) & -0.003(1) & 0.553(80) \\
3 & $V(q^2)$ & 16,19,20,23 & 0.8464(5) & 6.0 & 0.5072(35) & 0.069(2) & 1.101(123) \\
3 & $V(q^2)$ & 16,19,20,23 & 0.8464(5) & 8.39 & 0.5468(55) & 0.000(3) & 1.128(104) \\
\hline
\hline
\end{tabular}
\caption{The form factor results for $D_s \to \phi$ on each ensemble at all values of $q^2$ calculated. 
Column 2 denotes the form factor for the result in column 8. 
The axial form factors are fitted simultaneously so the (Goldstone) $D_s$ mass is the 
same for all of them on the same ensemble. The vector form factor $V(q^2)$ instead 
is calculated using the non-Goldstone $D_s$. 
Columns 5 and 6 give the value of $\theta$ used to give momentum to the $\phi$ 
and the fitted $\phi$ energy at this momentum.}
\label{tab:dsphidata}
\end{table*}

\subsubsection{Determining $A_1(q^2)$}
\label{subsub:a1}

From Eq.~\ref{eq:ffdef} we see that $A_1(q^2)$ is the 
only form factor that appears in the matrix element of 
the axial vector current when $\varepsilon^* \cdot q=0$, 
i.e. when the $\phi$ polarization is orthogonal to 
the momentum transfer. 
This is the only contribution to the matrix element at $q^2_{max}$ when the final state $\phi$ meson is at rest.
To calculate $A_1(q^2)$ away from $q^2_{max}$, we give the $\phi$ meson momentum in an orthogonal spatial direction to its polarization.

When the kinematics are set up such that $\varepsilon^* \cdot q = 0$, the transition matrix element becomes
\begin{equation}\langle \phi(p', \varepsilon) | A^\mu | D_s (p) \rangle = (m_{D_s}+m_\phi)\varepsilon^{*\mu} A_1(q^2). \label{eq:a1me} \end{equation}
The $\phi$ polarization vector and the axial vector current must then be 
in the same spatial direction so that the matrix element in 
Eq.~\ref{eq:a1me} is non-vanishing.

There is a choice of operators that can be used to extract $A_1(q^2)$: either using a local vector operator for the $\phi$ and local axial vector for the current or a 1-link vector and 1-link axial vector. 

If we use the 1-link $\phi$ with spin-taste $\gamma_\mu \otimes 1$ at time $0$, we also use the 1-link axial vector with spin-taste $\gamma_5\gamma_\mu \otimes \gamma_5$ at $t$ and the local pseudoscalar ($\gamma_5 \otimes \gamma_5$) for the $D_s$ at $T$ (see Fig. \ref{fig:3ptdiag}).
The staggered 3-point correlator is then
\begin{align}C_{3pt}(0,t,T) & = \sum_{x,y,z}(-1)^{x_\mu^<+ y_\mu^<}\varepsilon(x) \varepsilon(z) \label{eq:1linka1corr} \\
 & \times \mathrm{Tr}\left[g_s(x,z) g_c(z,y) g^{\theta\dag}_s(x\pm \hat{\mu},y \pm \hat{\mu})\right], \nonumber \end{align}
where the sites $x,y$ and $z$ are at times $0,t$ and $T$ and the sum is over the timeslice.
For the point-splitting at $x$ and $y$ we average over links in the forward and backward directions.
As the propagators are for staggered quarks, the trace is only over colour indices.

We write the staggered phase factors using
$\varepsilon(x) = \prod_\nu(-1)^{x_\nu}$, $x^<_\mu = \sum_{\nu<\mu} x_\nu$, $x^>_\mu = \sum_{\nu>\mu} x_\nu$ and $\bar{x}_\mu = \sum_{\nu\neq\mu} x_\nu$.

We fit the 3-point correlator simultaneously with the appropriate 2-point correlators.
In this case, these are the correlators for the Goldstone $D_s$ and 1-link $\phi$, given by
\begin{equation} C_{2pt, D_s}(0,t) = \sum_{x,y} \mathrm{Tr}\left[g_c(x,y) g^\dag_s(x,y)\right] \label{eq:goldstonecorr}\end{equation}
and
\begin{align} C_{2pt, \phi}(0,t) = & \sum_{x,y} \varepsilon(x)\varepsilon(y)(-1)^{x_\mu^<+y_\mu^<} \label{eq:1linkphicorr}\\
& \times \mathrm{Tr}\left[g_s(x,y) g^{\theta\dag}_s(x\pm\hat{\mu},y\pm\hat{\mu})\right]. \nonumber \end{align}
The sites $x$ and $y$ are again at times $0$ and $t$.
The results using these correlators are called `1-link $A_1(q^2)$' in Table \ref{tab:dsphidata}.

If we use local operators, we have a $\gamma_\mu \otimes \gamma_\mu$ operator at $0$ for the $\phi$, a $\gamma_\mu\gamma_5 \otimes \gamma_\mu\gamma_5$ axial vector at $t$ and the local pseudoscalar at $T$.
This gives the staggered 3-point correlation function
\begin{align}C_{3pt}(0,t,T) & = \sum_{x,y,z}(-1)^{x_\mu+ \bar{y}_\mu} \varepsilon(z) \\
 & \times \mathrm{Tr}\left[g_s(x,z) g_c(z,y) g^{\theta\dag}_s(x,y)\right]. \nonumber \end{align}
The corresponding $D_s$ 2-point correlator is that given by Eq.~\ref{eq:goldstonecorr} and the local $\phi$ correlator is
\begin{align} C_{2pt, \phi}(0,t) = & \sum_{x,y} (-1)^{x_\mu+y_\mu} \label{eq:localphicorr}\\
& \times \mathrm{Tr}\left[g_s(x,y) g^{\theta\dag}_s(x,y)\right]. \nonumber \end{align}
The determination of $A_1(q^2)$ using a local axial current is called `Local $A_1(q^2)$' in Table \ref{tab:dsphidata}.
The two determinations of $A_1(q^2)$ should agree with one another, 
as the current operators and $\phi$ mesons only differ by taste. 
Doing the calculation in two different ways allows us to test the differences between different tastes of staggered mesons in meson transitions.
Agreement is seen in the results for $A_1(q^2)$ given in Table \ref{tab:dsphidata} within 
our statistical errors.
We find smaller statistical errors for Local $A_1(q^2)$ on the coarse lattices (sets 1 and 2), 
so it is used for the final results.

\subsubsection{Determining $A_0(q^2)$}
\label{subsub:a0}
The $A_0(q^2)$ form factor can be related to the pseudoscalar density using the PCAC relation so can be extracted from the pseudoscalar matrix element using Eq.~\ref{eq:pme}.

The staggered 3-point correlator calculated on the lattice then has $\gamma_5 \otimes \gamma_5$ at both $t$ and $T$ and a 1-link $\phi$ with spin-taste $\gamma_\mu \otimes 1$ at $0$.
The correlator is
\begin{align}C_{3pt}(0,t,T) & = i\sum_{x,y,z}(-1)^{x_\mu^<} \varepsilon(x)\varepsilon(y)\varepsilon(z) \label{eq:a0corr} \\
 & \times \mathrm{Tr}\left[g_s(x,z) g_c(z,y) g^{\theta\dag}_s(x\pm \hat{\mu},y)\right]. \nonumber \end{align}
It cannot be calculated at rest because $\varepsilon^* \cdot q$ must be non-zero for the matrix element to be non-zero.
The $\phi$ meson therefore carries momentum in the same direction as its polarization.
We fit the 3-point correlators generated using Eq.~\ref{eq:a0corr} simultaneously 
with $D_s$ and $\phi$ 2-point correlators as given in Eqs.~\ref{eq:goldstonecorr} and~\ref{eq:1linkphicorr}.
The results are given in Table \ref{tab:dsphidata}.

\subsubsection{Determining $A_2(q^2)$}
\label{subsub:a2}
The form factor $A_2(q^2)$ is difficult to calculate as it only contributes to the matrix element when $\varepsilon^* \cdot q \neq 0$.
This requires the $\phi$ meson's polarization, $\varepsilon^*$, and momentum, $p'$, to be in the same direction.
In this case, all of the axial form factors, $A_1(q^2)$, $A_2(q^2)$ and $A_0(q^2)$, appear and we calculate $A_2(q^2)$ given $A_1(q^2)$ and $A_0(q^2)$.

At $q^2 = 0$, we have a relationship between $A_0$, $A_1$ and $A_2$ because $A_3$ is 
given in terms of $A_1$ and $A_2$ by Eq.~\ref{eq:a3def} and $A_3(0) = A_0(0)$.
This means we can extract $A_2(0)$ from the values of $A_1(0)$ and $A_0(0)$.

Away from $q^2 = 0$, we no longer have the relation between $A_0(q^2)$ and $A_3(q^2)$ so extracting $A_2(q^2)$ is more complicated.
If we calculate the 3-point correlation function given by Eq.~\ref{eq:1linka1corr}, but with the $\phi$ polarization, $\varepsilon^*$, parallel to its momentum, $p'$, the result depends of all the axial form factors, and is given by Eq.~\ref{eq:ffdef}.
Using the values of $A_1(q^2)$ and $A_0(q^2)$ determined as described in Sections \ref{subsub:a1} and \ref{subsub:a0}, $A_2(q^2)$ can be extracted.
The results given in Table \ref{tab:dsphidata} use the local determination of $A_1(q^2)$ to extract $A_2(q^2)$, but agree with using 1-link $A_1(q^2)$.

\subsubsection{Determining $V(q^2)$}
\label{subsub:v}

The vector form factor, $V(q^2)$, is the same form factor that appears in electromagnetic vector to pseudoscalar meson transitions, such as $J/\psi \to \eta_c \gamma$ \cite{jpsi}.
The form factor can be calculated with staggered quarks using the same 3-point correlator setup we used for $J/\psi \to \eta_c \gamma$.

We calculate $V(q^2)$ using the non-Goldstone $D_s$ and a 1-link vector operator for the $\phi$ where the point-splitting is in a different spatial direction to the polarization.
In spin-taste notation, this operator is $\gamma_\mu \otimes \gamma_\mu\gamma_\nu$ and it is placed at time $0$.
As the $D_s$ meson is at rest, the $\phi$ must carry non-zero momentum.
The non-Goldstone $D_s$ at time $T$ is simulated using a $\gamma_5\gamma_t \otimes \gamma_5\gamma_t$ operator and there is a local $\gamma_\alpha \otimes \gamma_\alpha$ vector operator at $t$. 
The 3-point correlation function is
\begin{align}C_{3pt}(0,t,T) & = i\sum_{x,y,z}(-1)^{x_\nu^> + x_\mu+ y_\alpha +z_t}  \\
& \times \mathrm{Tr}\left[g_s(x,z) g_c(z,y) g^{\theta\dag}_s(x\pm \hat{\nu},y)\right]. \nonumber \end{align}
The corresponding non-Goldstone $D_s$ 2-point function is given by
\begin{equation} C_{2pt, D_s}(0,t) = \sum_{x,y} (-1)^{\bar{x}_t+\bar{y}_t} \times \mathrm{Tr}\left[g_c(x,y) g^\dag_s(x,y)\right] \label{eq:nongoldstonecorr}\end{equation}
and the $\phi$ 2-point correlator by
\begin{align} C_{2pt, \phi}(0,t) = & \sum_{x,y} (-1)^{x_\nu^>+x_\mu+y_\nu^>+y_\mu} \\
& \times \mathrm{Tr}\left[g_s(x,y) g^{\theta\dag}_s(x\pm\hat{\nu},y\pm\hat{\nu})\right]. \nonumber \end{align}

As the 3-point correlator must be a taste-singlet overall, $\mu, \nu$ and $\alpha$ must be three different spatial directions.
The $\phi$ meson polarization vector, vector current and $\phi$ momentum must also all be orthogonal to one another due to the $\epsilon^{\mu\nu\alpha\beta}$ in Eq.~\ref{eq:ffdef}.
Therefore the momentum of the $\phi$ must be in the $\nu$ direction. 
For the vector form factor, Eq.~\ref{eq:ffdef} reduces to
\begin{equation} \langle \phi(p^{\prime},\varepsilon) | V_{\alpha} | D_s(p) \rangle = \frac{2i\epsilon_{\alpha\mu\nu t}}{M_{D_s}+M_{\phi}} \varepsilon^{*\mu} p'^\nu p^t V(q^2). \label{eq:vff} \end{equation}
As the $D_s$ meson is at rest, only the time component of its 4-momentum is non-zero so Equation \ref{eq:vff} must include $p_t$, the $D_s$ energy (in this case, mass).
We give our results for $V(q^2)$ in Table \ref{tab:dsphidata}.

\begin{figure}
\centering
\includegraphics[angle=-90,width=0.45\textwidth]{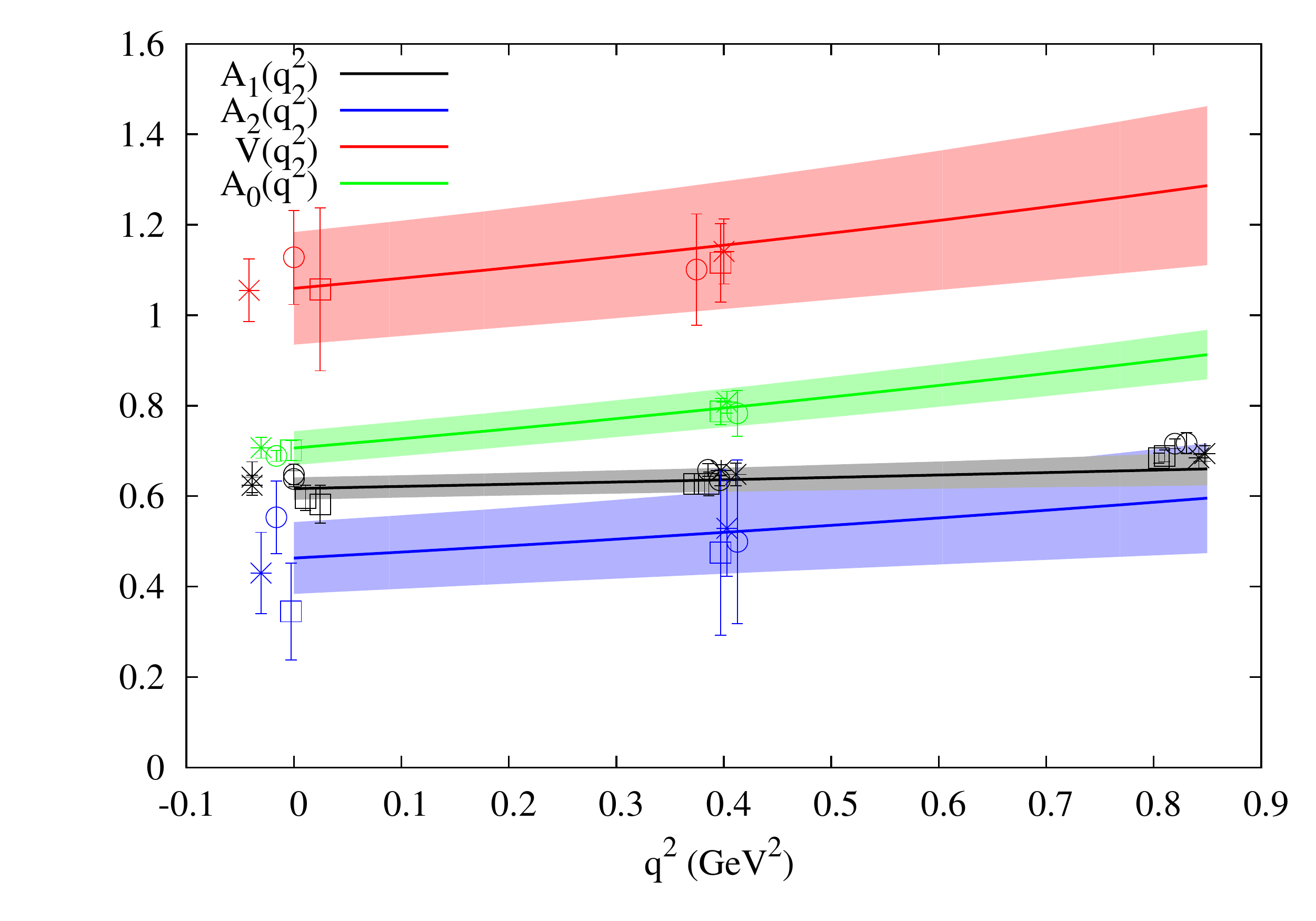}
\caption{Lines and bands show the form factors from our lattice QCD calculation extrapolated 
to the physical point as a function of $q^2$ for $q^2=0$ to $q^2_{max}$. The form factor $A_1(q^2)$ is shown in black, $A_2(q^2)$ in blue, $V(q^2)$ in red and $A_0(q^2)$ in green. 
We also plot the raw lattice results from Table~\ref{tab:dsphidata} for each form factor. The bursts are the lattice data from Set 1, squares from Set 2 and circles Set 3. At $q^2_{max}$, the $\phi$ meson is at rest and only $A_1(q^2)$ contributes to the decay. We plot the lattice data for both the local and 1-link determinations of $A_1(q^2)$.}
\label{fig:formfactors}
\end{figure}

\subsubsection{Comparing form factors to experiment}
\label{subsub:fexpt}
We fit the correlators for all the axial form factors at all values of $q^2$ simultaneously on each ensemble. 
This allows us to use the correlation matrix between them in our physical extrapolation.
The results in Table \ref{tab:dsphidata} are taken from fits with 5 exponentials for Sets 1 and 2 and 4 exponentials for Set 3. 

The extrapolation of the form factors to zero lattice spacing and physical light quark masses is more complicated than for meson masses and decay constants since we want 
to determine the functional form as a function of $q^2$.
It is convenient to parameterise the form factors using the z-expansion~\cite{arnesenzexp, hillzexp, lellouchzexp}.
The conversion from $q^2$ to $z$ is made using the transformation
\begin{equation} z(q^2) = \frac{\sqrt{t_+ - q^2} - \sqrt{t_+ - t_0} }{\sqrt{t_+ - q^2} + \sqrt{t_+-t_0} } , \label{eq:q2toz} \end{equation}
where $t_+ = (m_{D_s}+m_\phi)^2$.
We will use $t_0 = 0$ here which means that $z=0$ corresponds to $q^2=0$. 
The z-expansion maps the line above the real axis from $q^2=\infty$ to $q^2=t_+$ and then 
back below the real axis to $\infty$ onto the unit circle. The semileptonic 
region $0< q^2 < q^2_{max}=t_-=(m_{D_s}-m_{\phi})^2$ is then mapped to a line 
inside this circle. The shape of the form factor can be described by a power series in $z$.
As $z$ is small, the series can be truncated and the form factor described with only a few terms.

Physical particles of $c\overline{s}$ quark content 
that have appropriate quantum numbers for that form 
factor and masses between $t_+$ and $t_-$ will appear 
as poles inside the unit circle. We therefore remove those before 
we transform to $z$-space:
\begin{equation} \tilde{A}_i(q^2) = \left( 1 - \frac{q^2}{M_{D_{s1}}^2} \right) A_i(q^2) \end{equation}
and 
\begin{equation} \tilde{V}(q^2) = \left( 1 - \frac{q^2}{M_{D_s^*}^2} \right) V(q^2).\end{equation}
The pole masses are $M_{D_s^*} = 2112$ MeV for the vector and $M_{D_{s1}} = 2459$ MeV for the axial vector \cite{pdg}. The pole factors here are relatively benign because 
$\sqrt{q^2_{max}}$ for $D_s \rightarrow \phi$ decay is much smaller than either of these 
masses. This is also the reason why we only divide out one pole in each case, and do not 
consider higher mass particles.  
The z-expansion can then be used to extrapolate the lattice QCD form factors 
to the physical limit by making the coefficients of the terms in z-space depend on the lattice spacing and sea quark masses \cite{Na:2010uf, jonnadtok}.

For each of the form factors, $\tilde{F}(z)$, we use the fit function
\begin{equation} \tilde{F}(z) = \sum^3_{n=0} B^F_n\left\{ 1 + C^F_na^2 + D^F_na^4 + E^F_n x_l \right\} z^n. \end{equation}
The fit parameter $B^F_0$ is the form factor at $z=q^2=0$ and chiral parameter $x_l = m_l/m_{s, phys}$ is given in Table \ref{tab:params}.
We include up to $n=3$ in our fit as we find that all higher terms make no difference to the results.

All the form factors are fitted together to this form, but the coefficients are independent for each one.
The priors are taken as $0.0(2.0)$ for $B^F_n (n>0)$ and  $0.0(1.0)$ for $C^F_n$, $D^F_n$ and $E^F_n$.
The priors for the form factors at $z=0$ are $B^{A_1}_0 = 0.6(0.2)$, $B^{A_2}_0 = 0.4(0.2)$, $B^{V}_0 = 1.0(0.2)$ and $B^{A_0}_0 = 0.7(0.2)$.
As the values of the form factors at $q^2=0$ are given by fit parameters, we can enforce the kinematic constraint $A_3(0)=A_0(0)$ by replacing $B_0^{A_0}$ with $\frac{M_{D_s} + M_\phi}{2M_\phi}B_0^{A_1} - \frac{M_{D_s} - M_\phi}{2M_\phi} B_0^{A_2}$.
This does not significantly alter the results obtained from our fit, since they are consistent 
with the constraint without imposing it.
The physical $z$-expansion of the form factor is obtained by setting $a=0$ and 
$x_l = 1/27.5$~\cite{pdg}. 

\begin{table} 
\begin{center} 
\begin{tabular}{c|c} 
\hline 
\hline 
Form factor & Ratio \\
\hline 
$A_1(0)$ = $0.615(24)$ & -- \\
$A_2(0)$ = $0.457(78)$ & $r_2$ = $0.74(12)$ \\
$A_0(0)$ = $0.706(37)$ & $r_0$ = $1.14(6)$\\
$V(0)$ = $1.059(124)$ & $r_V$ = $1.72(21)$\\
\hline 
\hline 
\end{tabular} 
\caption{The form factors calculated on the lattice at maximum recoil. For all the form factors other than $A_1$, we also give the ratio of the form factor at $q^2=0$ to $A_1(0)$. 
}
\label{tab:ffq20} 
\end{center} 
\end{table}

After the extrapolation, the form factors in the physical limit are converted back to $q^2$ space.
We plot the $A_1(q^2)$, $A_0(q^2)$, $A_2(q^2)$ and $V(q^2)$ form factors against $q^2$ 
for the full physical range of $q^2$ values in Figure~\ref{fig:formfactors}. 
The solid lines are the central values of the form factors after the extrapolation and the shaded bands show the errors. 
The raw lattice results for each of these form factors are also plotted with symbols 
in Figure~\ref{fig:formfactors}. 
We see that the form factors agree well on each set of gauge configurations and 
do not vary significantly with lattice spacing or sea quark masses, so the extrapolation 
to the physical point changes the results very little.

Our results are most accurate for the $A_1$ form factor. In Table~\ref{tab:ffq20} 
we give our values for each form factor at $q^2=0$ and its ratio to the 
$A_1$ form factor at that point. 
The ratios can be compared to experimental results from BaBar~\cite{babar} who quote 
$r_V = V(0)/A_1(0) = 1.849(60)(95)$ and $r_2 = A_2(0)/A_1(0) = 0.763(71)(65)$.
We find $r_V = 1.72(21)$ and $r_2 = 0.74(12)$, in agreement with experiment.
We can also take the ratio $r_0 = A_0(0)/A_1(0)$ on the lattice and we find $r_0 = 1.14(6)$.
The error in the lattice QCD results is dominated in all cases by the 
statistical error in the raw lattice results. 

\begin{figure}
\centering
\includegraphics[angle=-90,width=0.45\textwidth]{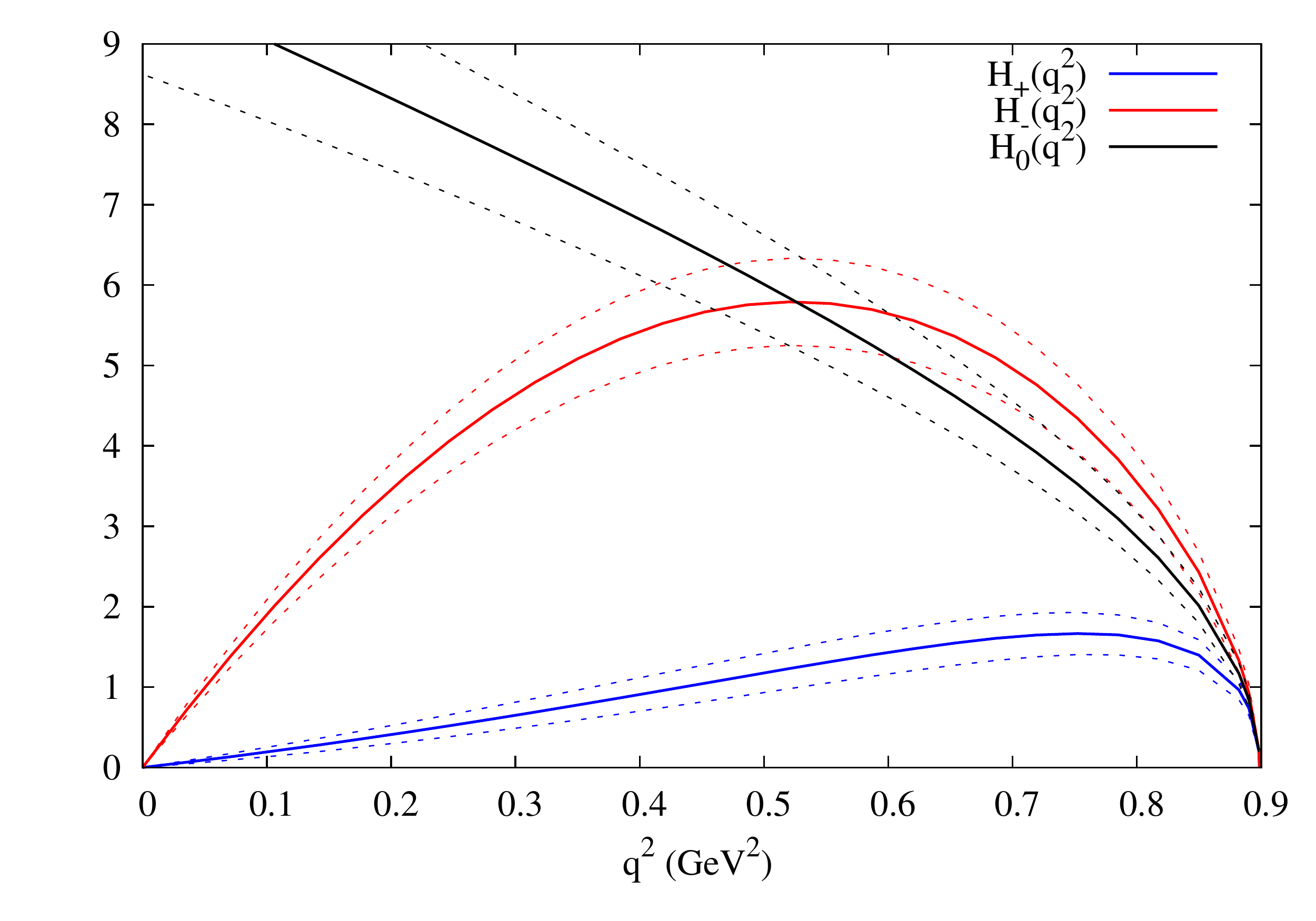}
\caption{The $D_s \to \phi \ell \nu$ helicity amplitudes constructed from our extrapolated form factors. They are shown as $p_\phi q^2 |H_\pm(q^2)|^2$ and $p_\phi q^2 |H_0(q^2)|^2$, including the kinematic factors that appear in the differential decay rate. At the $q^2=q^2_{max}$ end of the distribution, $p_\phi \to 0$ and each of the helicity amplitudes vanishes.}
\label{fig:helicityamps}
\end{figure}

To extract the differential decay rate from Eq.~\ref{eq:diffrate}, we need to 
combine the form factors into the helicity amplitudes given in Eqs.~\ref{eq:hpm} and~\ref{eq:h0}.
The helicity amplitudes appear in Eq.~\ref{eq:diffrate} as $p_\phi q^2 |H_i(q^2)|^2$, so we plot this combination as a function of $q^2$ in Figure \ref{fig:helicityamps} for 
$H_\pm(q^2)$ and $H_0(q^2)$.
In Figure \ref{fig:helicityamps}, we include a multiplying 
factor of $\frac{32}{9}$ from the angular integration; the factor is the same for each helicity. 
This means we are plotting the contribution to the differential decay 
rate as a function of $q^2$ for each helicity.
The cross terms between different helicity amplitudes in Eq.~\ref{eq:diffrate} vanish when 
we integrate over the angle $\chi$ so these only affect the distribution in $\chi$, and not in $q^2$.
At low $q^2$, the decay rate is dominated by $H_0(q^2)$.
Throughout the range of $q^2$, both $H_0(q^2)$ and $H_-(q^2)$ contribute more than $H_+(q^2)$, 
as expected from the $V-A$ nature of the weak interaction. 
As we plot the combination $p_\phi q^2 |H_i(q^2)|^2$, all the helicity amplitudes go to zero at $q^2_{max}$ because $p_\phi=0$ in this limit.

\begin{figure}
\centering
\includegraphics[angle=-90,width=0.45\textwidth]{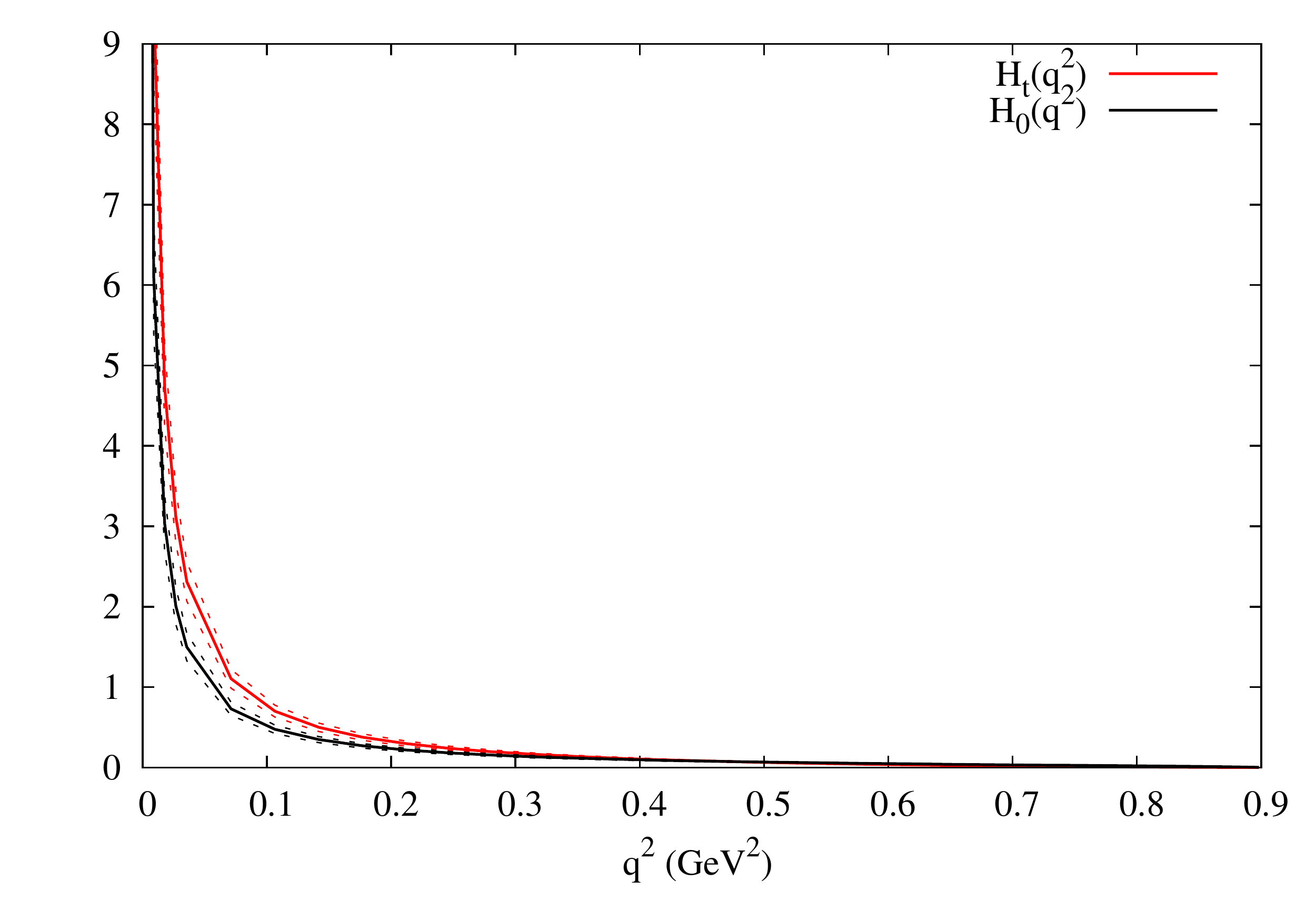}
\caption{Helicity amplitude contributions to $D_s \to \phi \ell \nu$ which are suppressed by the lepton mass. 
Here we plot the contributions from $H_0(q^2)$ and $H_t(q^2)$ which are enhanced at small 
values of $q^2$. They are plotted as they appear in the differential decay rate for 
muons, i.e. $m_\mu^2 p_\phi |H_i(q^2)|^2$. The scale is the same as in 
Figure \ref{fig:helicityamps}. For semileptonic decay with electrons in the final 
state, these contributions are further suppressed.}
\label{fig:suppressedhelicityamps}
\end{figure}

The terms which are suppressed by the lepton mass appear in the differential 
decay rate as $m_\ell^2 p_\phi |H_i(q^2)|^2$ and the largest contributions come from $H_t(q^2)$ and $H_0(q^2)$ which are enhanced at low values of $q^2$.
In Figure~\ref{fig:suppressedhelicityamps}, we plot these contributions 
as $m_\mu^2 p_\phi |H_i(q^2)|^2$, again including factors from the angular integrals, 
so that these can be compared directly with Figure~\ref{fig:helicityamps}.
The scale is the same and shows that these contributions are only large at small values of $q^2$. 
The decay $D_s \rightarrow \phi \mu \nu$ has been studied by FOCUS~\cite{Link:2004qt}. 
Here we compare our results in most detail to those of BaBar~\cite{babar} who measured 
the rate for $D_s \rightarrow \phi e \nu$. 
For electrons in the final state, the contributions shown in Figure~\ref{fig:suppressedhelicityamps} 
are smaller by a factor of $m_e^2 / m_\mu^2$ (=$2\times10^{-5}$) 
and will not be visible.

\subsubsection{Determining $V_{cs}$}
\label{subsub:vcs}
\begin{figure*}
\centering
\includegraphics[angle=-90,width=0.90\textwidth]{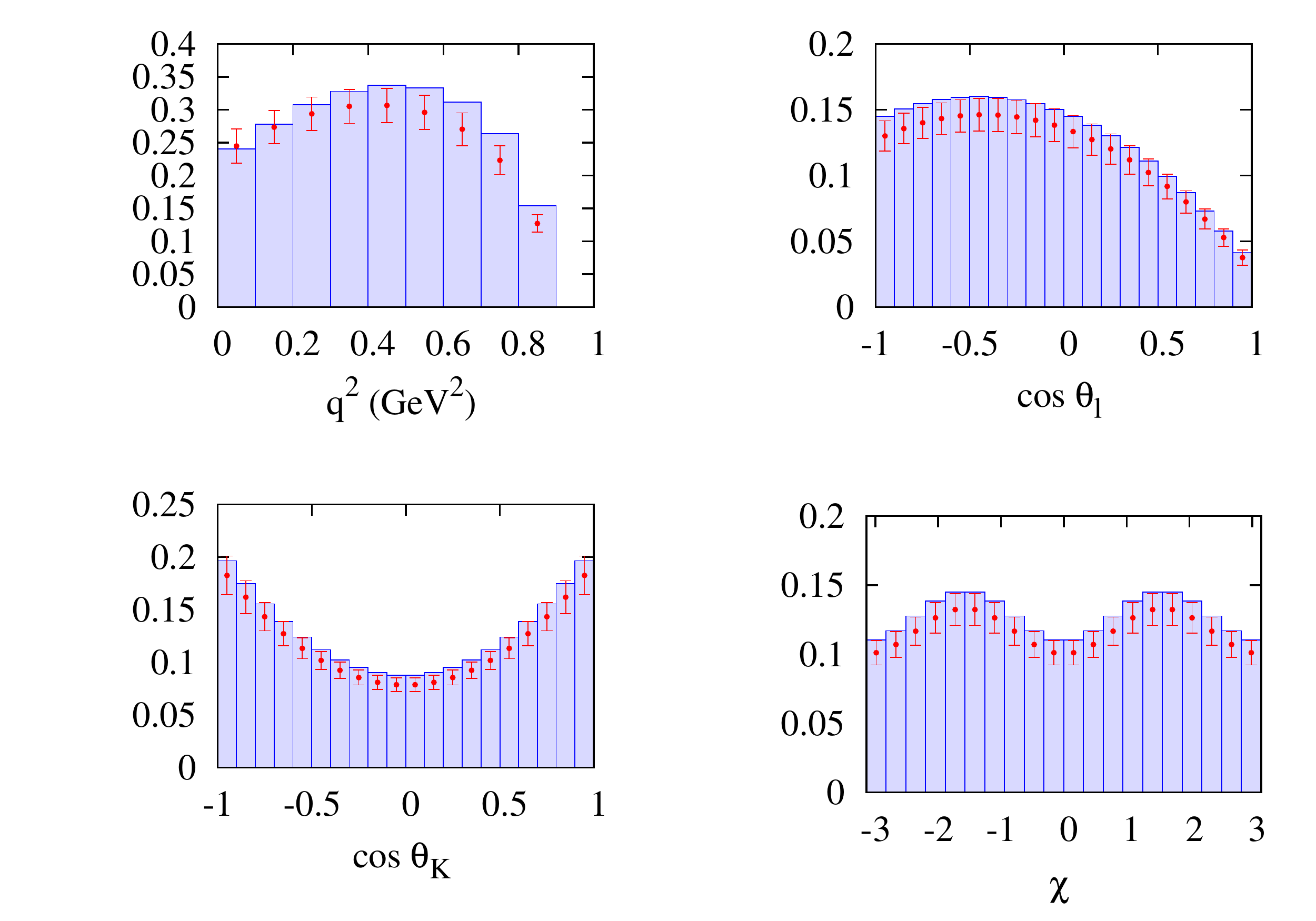}
\caption{Decay distributions for $D_s \to \phi \ell \nu$ with $\phi \to K^+ K^-$ for each of the kinematic variables in the decay. The decay angles are shown in Figure \ref{fig:angles} and described in the text. The lattice results are shown with the red points with error bars and the experimental results by blue blocks. The experimental errors are not plotted, but are of similar size to the lattice errors.}
\label{fig:dists}
\end{figure*}

The differential decay rate for $D_s \rightarrow \phi \ell \nu$ is given 
by Eq.~\ref{eq:diffrate} and we can plot it as a function of each of $q^2$, $\cos \theta_K$, $\cos \theta_\ell$ and $\chi$ by integrating over the other three.
The angular integrals are straightforward and we integrate over $q^2$ numerically.
The distributions we obtain from our form factors are plotted in Figure~\ref{fig:dists}, 
where we take the value of $V_{cs}$ from unitarity and 
use $\mathcal{B}r(\phi \to K^+K^-) = 0.489(6)$ \cite{pdg}.
Our lattice results are plotted as red data points with errors and 
the experimental results from BaBar~\cite{babar} are plotted as the blue histogram.
To avoid the effects of experimental cuts on the distributions (particularly on the lepton momentum), we reconstruct the decay rate in each bin from the results quoted by BaBar 
for the ratios of form factors at $q^2 =0$ and the pole masses for the $q^2$ distributions 
they obtain from fits to their data. 
The experimental errors are not plotted, but they are of a similar size to our lattice errors.
There is good agreement between the lattice results and experiment both 
in magnitude and shape for each of the differential distributions. 
We discuss the different distributions one by one below. 

After performing the angular integrals, all of the helicity factors, $|H_+(q^2)|^2$, $|H_-(q^2)|^2$ and $|H_0(q^2)|^2$ have the same coefficient in the $q^2$ distribution.
The relative contribution of each one as a function of $q^2$ is then as 
shown by Figure \ref{fig:helicityamps}. 


The $\cos\theta_\ell$ distribution is dominated by $H_0$ at $\cos\theta_\ell = 0$.
$|H_0|^2$ appears with a factor of $(1-\cos^2\theta_\ell)$ so makes no 
contribution at the $\cos\theta_\ell = \pm1$ ends of the distribution.
At $\cos\theta_\ell = 1$, the only helicity that contributes is $H_+$ and at $\cos\theta_\ell = -1$, only $H_-$ contributes as these helicity amplitudes appear with factors $(1\pm\cos\theta_\ell)^2$.
We see that the distribution is larger at $\cos\theta_\ell = -1$ than $+1$, which is a result of the dominance of $H_-$ over  $H_+$ coming from the $V-A$ weak interation. 


The helicities $H_\pm$ both contribute to the $\cos\theta_K$ distribution as $1-\cos^2\theta_K$ and dominate at $\cos\theta_K=0$.
At $\cos\theta_K =\pm 1$, the only contribution is from $H_0$, which contributes as $\cos^2\theta_K$.
The coefficients from the integrals over $\cos\theta_\ell$ and $\chi$ are $\frac{16}{3}$ for $|H_0|^2$ and $\frac{8}{3}$ for $|H_+|^2+|H_-|^2$.


The $\chi$ distribution is a constant with an oscillation of $-\cos2\chi H_+(q^2)H_-(q^2)$ as the $H_0(q^2)H_\pm(q^2)$ terms 
in Eq.~\ref{eq:diffrate} vanish when we integrate over the other angles. 

By integrating over all of the kinematic variables, we can calculate the total decay rate. 
We can then extract $V_{cs}$ by comparing the total decay rate to that 
measured by BaBar in~\cite{babar}.
We take BaBar's branching ratio for $D_s \to \phi e^+ \nu_e$ of $2.61(17)\times 10^{-2}$ and $\tau_{D_s} = 500(7) \times 10^{-15}s$~\cite{pdg}.
The experimental measurements and lattice calculation differ by a factor of $|V_{cs}|^2$ 
so we obtain $V_{cs} = 1.017(44)_{\mathrm{latt}}(35)_{\mathrm{expt}}(30)_{K\overline{K}}$.
The error from lattice QCD includes statistical errors from the lattice data (which dominate), 
uncertainty in the determination of the weak current $Z$ factors and the extrapolation to 
the physical point. 
The final error takes into account the fact that the $\phi$ meson has 
a strong decay mode to $K\bar{K}$.
As discussed in Section \ref{subsec:phi}, we estimate this error to be 3\%.
This gives us a final result of $V_{cs} = 1.017(63)$.

This value is in agreement with unitarity~\cite{pdg} and other lattice 
measurements of $V_{cs}$ from $D_s$ leptonic~\cite{Dsdecayconst} decay and 
$D \rightarrow K \ell \nu$ 
semileptonic~\cite{jonnadtok} decay.

\section{Discussion}
\label{sec:discussion}

For pseudoscalar to pseudoscalar meson transitions, we have found that the form factors agree for $D \to K$ and $D_s \to \eta_s$ to within 2\%~\cite{jonnadtok}.
These decays differ only by whether the spectator quark in the 
decay is a light or strange quark.
To test whether the same is true for the pseudoscalar to vector transitions, we 
can compare the form factors we extract for $D_s \to \phi$ with 
experimental results for $D \to K^*$. 

CLEO~\cite{cleodkstar} give their reconstructed values of $q^2 |H_i(q^2)|^2$ for $i=\pm,0$ in $q^2$ bins
We construct the $D \to K^*$ helicity amplitudes from Eqs.~\ref{eq:hpm} and \ref{eq:h0}, using the same form factors as $D_s \to \phi$ and replacing the meson masses and kinematic factors with those appropriate for $D \to K^*$.

\begin{figure}
\hskip -10pt
\includegraphics[angle=-90,width=0.45\textwidth]{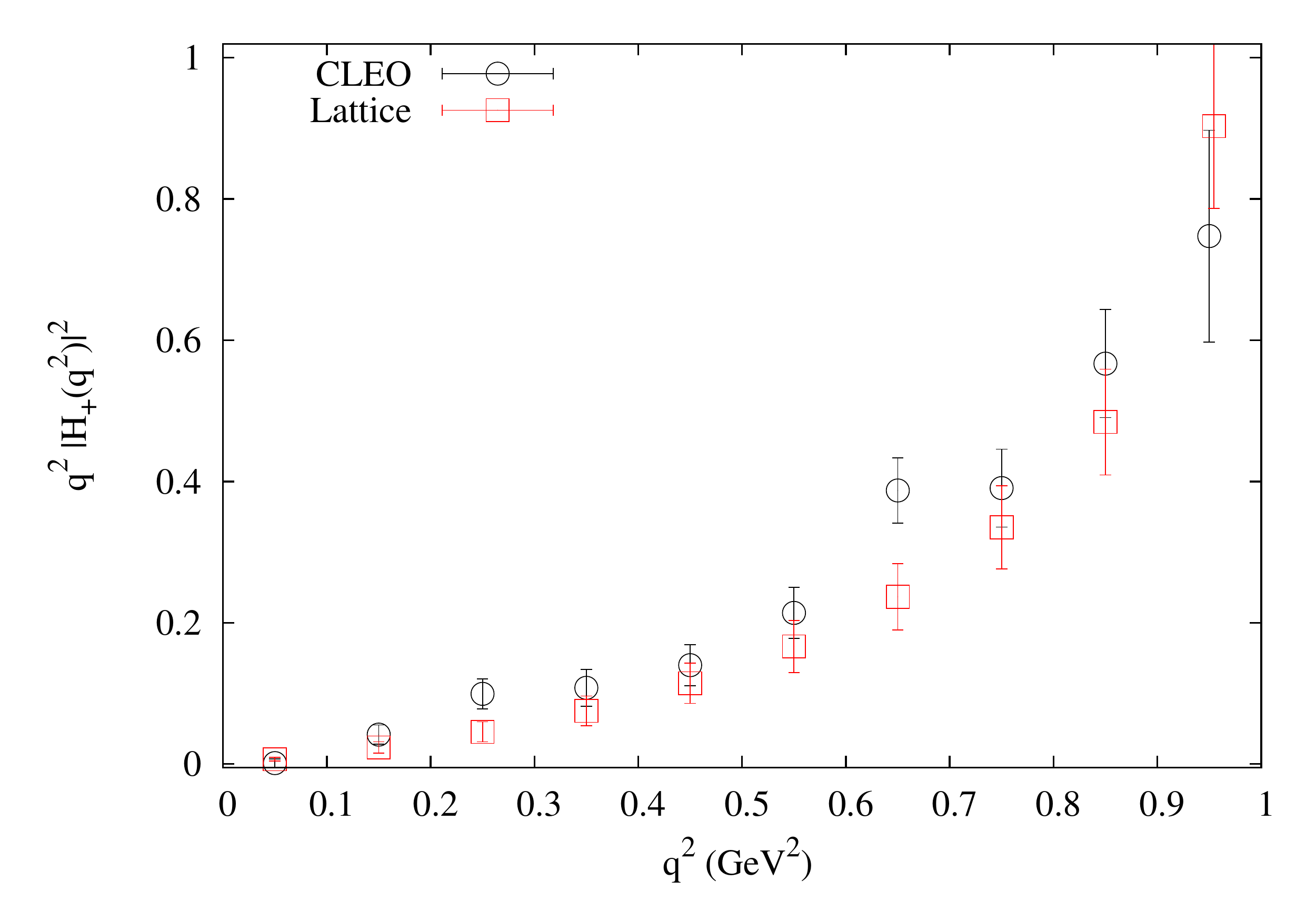} \\[1em]
\hskip -10pt
\includegraphics[angle=-90,width=0.45\textwidth]{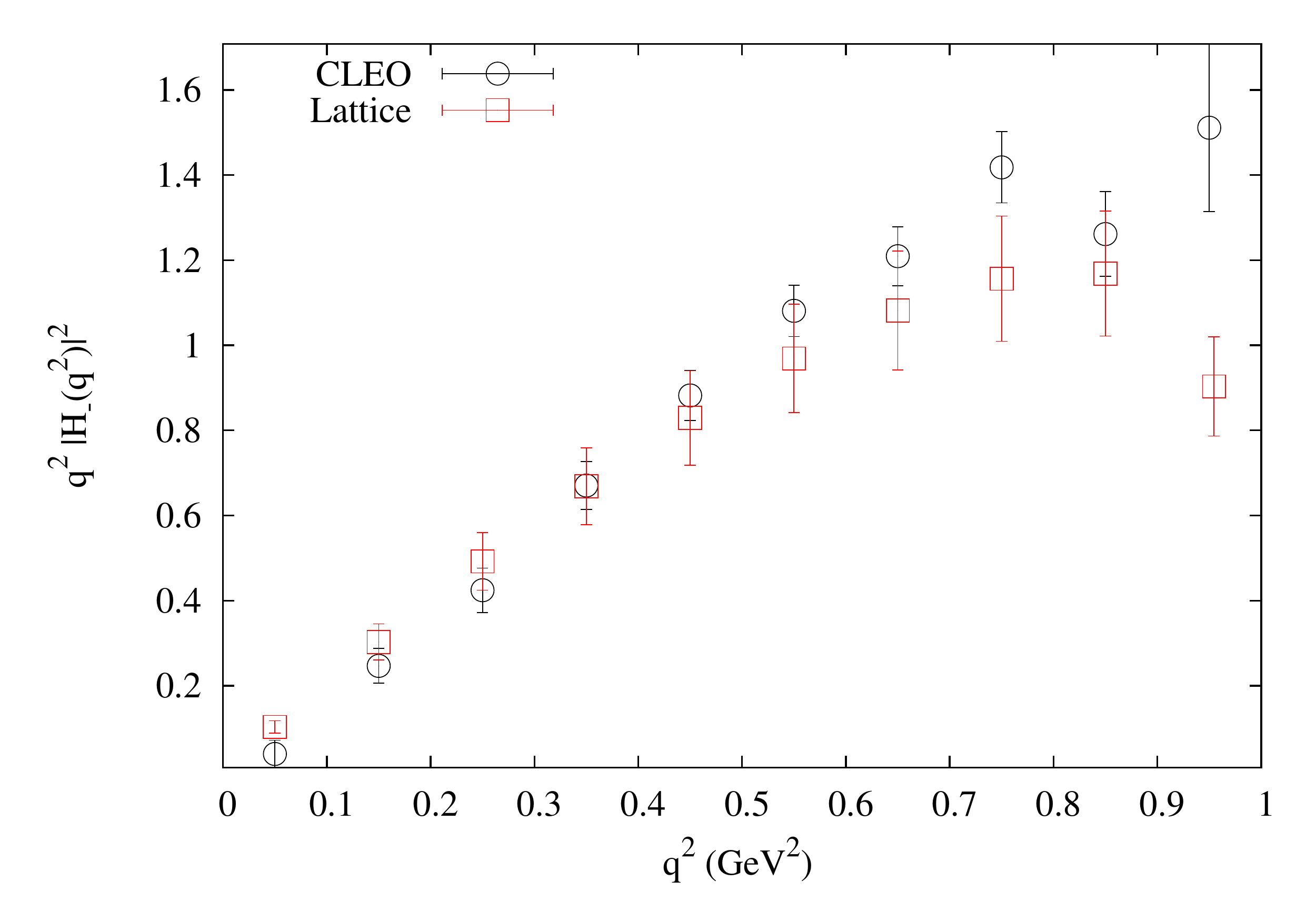} \\[1em]
\hskip -10pt
\includegraphics[angle=-90,width=0.45\textwidth]{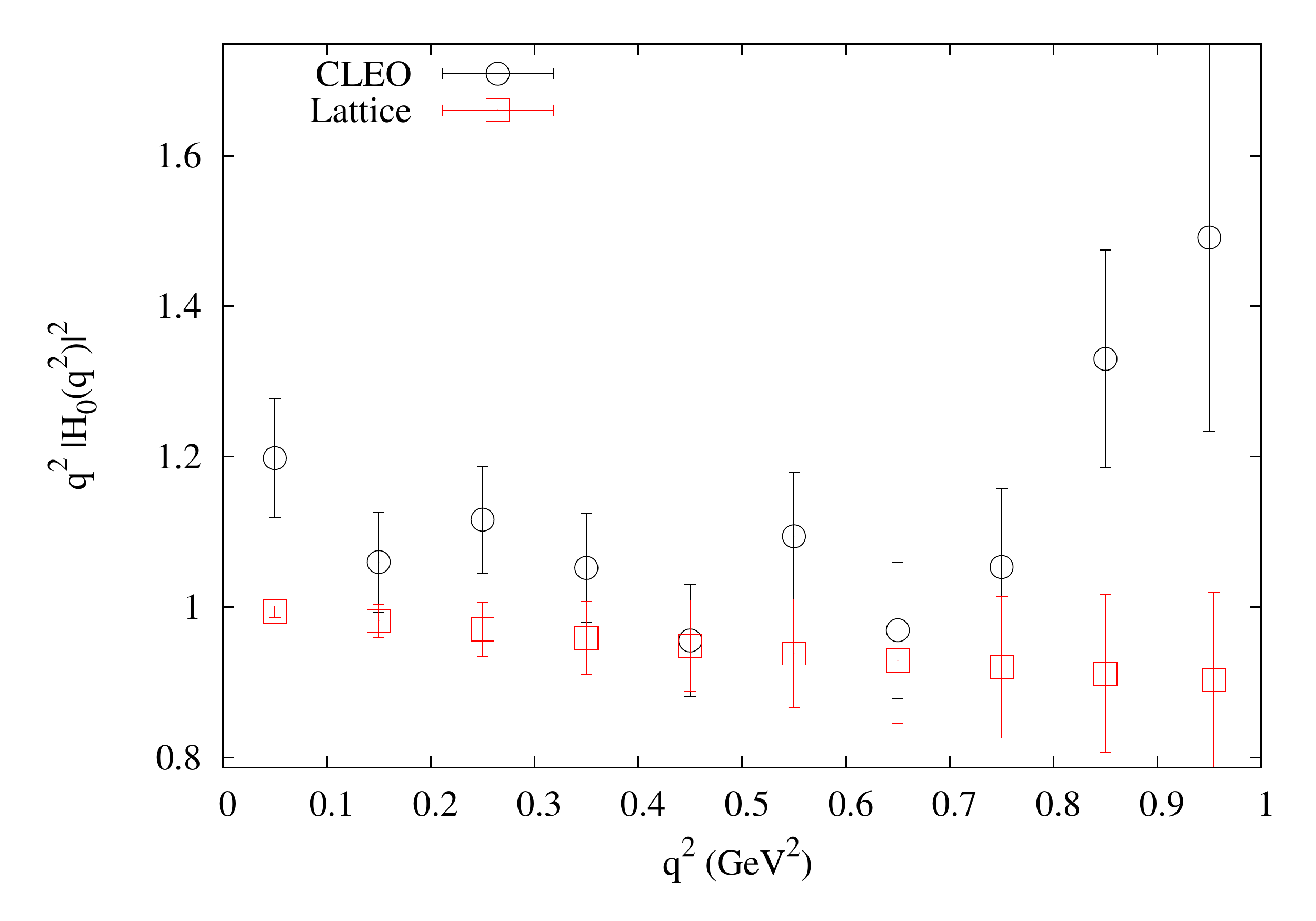} 
\caption{Comparison of lattice QCD with CLEO's determination~\cite{cleodkstar} of the helicity amplitudes for $D \to K^*$. We assume that the form factors are insensitive to the spectator quark mass, so we can construct the $D \to K^*$ helicity amplitudes using the same form factors as $D_s \to \phi$. The data plotted is normalised such that $q^2 |H_0(q^2)|^2 \to 1$ as $q^2 \to 0$.}
\label{fig:cleo}
\end{figure}

In Figure \ref{fig:cleo}, we plot $q^2 H_i^2(q^2)$ for $H_\pm$ and $H_0$. 
The lattice data is plotted in red and CLEO's results are in black.
The final red point is offset slightly from experiment -- it is at $q^2_{max} = 0.954~\mbox{GeV}^2$. 
The CLEO data is normalised by $q^2 |H_0(q^2)|^2 \to 1$ as $q^2 \to 0$, so we apply 
the same normalisation condition to the lattice results.

There is reasonable agreement between lattice and experiment, which indicates that the semileptonic form factors for $D_s \to \phi$ and $D \to K^*$ also show little dependence on the spectator quark mass.
However, the comparison between the $D_s \to \phi$ and $D \to K^*$ decays may be further complicated by the vector particles' widths.
We have treated the $\phi$ meson as stable in our calculation of the semileptonic form factors and take a systematic error, as described in Section \ref{subsec:phi}, to account for this.
The width of the $K^*$ is considerably larger than that of the $\phi$ and this may have a larger effect on the $D \to K^* \ell \nu$ decay, limiting the extent to which we can expect it to be 
well described by the $D_s \to \phi \ell \nu$ form factors we have calculated.
This strong decay mode also makes it difficult to calculate $D \rightarrow K^*$ form factors 
directly to high accuracy in lattice QCD. 

We do not compare the results for $H_t(q^2)$ because the experimental errors are too large. 
However, CLEO are able to use the ratio of semileptonic decays to electrons and muons to 
extract information about the lepton-mass-suppressed helicity functions. 
This is encouraging for a future comparison of these helicity functions between 
lattice QCD and experiment.  

\begin{figure}
\includegraphics[angle=-90,width=0.48\textwidth]{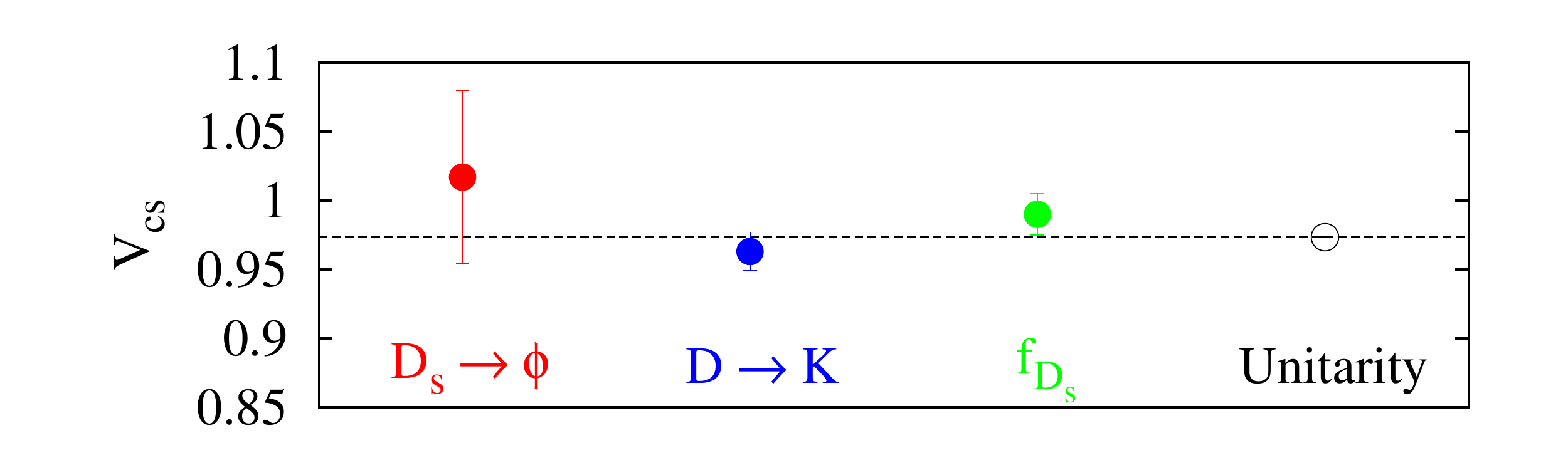} 
\caption{Comparison of the values of $V_{cs}$ obtained here from $D_s \rightarrow \phi$ 
decay and those from $D_s$ leptonic decay~\cite{Dsdecayconst} and $D \rightarrow K$ semileptonic decay~\cite{jonnadtok}. 
We also show the result from CKM unitarity~\cite{pdg}. 
}
\label{fig:vcs}
\end{figure}

\section{Conclusions} 
\label{sec:conclusions}

We have calculated the complete set of axial vector and vector form factors for 
a pseudoscalar to vector weak semileptonic decay from full lattice QCD 
for the first time. 
We chose the process $D_s \to \phi \ell \nu$ because the initial and final 
mesons contain no valence light quarks and so we can do a relatively accurate 
calculation. 
We are also able to cover the full range of $q^2$ available to the decay.  
Calculating all of the form factors allows us to construct the angular 
decay distributions which contain information about the helicity of the 
$W$ boson because the final state particle is a vector. 
The distributions we obtain are in good agreement with those observed by experiment.

Comparison of the total rate, integrated over all kinematic variables, with the 
experimentally measured branching fraction allows us to extract a value for the CKM element $V_{cs}$.
Our final result for $V_{cs}$ is $1.017(44)_{\mbox{latt}}(35)_{\mbox{expt}}(30)_{K\bar{K}}$.
In Figure~\ref{fig:vcs}, we compare this value with those from the lattice determination of 
the $D_s$ decay constant and its comparison to $D_s$ leptonic decay and from the $D \rightarrow K$ 
form factor and its comparison to $D$ semileptonic decay. The value from 
CKM unitarity is also shown. 
We see that the result from $D_s \rightarrow \phi$ decay is in good agreement with the 
other results but has 
larger errors. This is a combination of both larger lattice QCD and experimental errors.
The lattice QCD error is dominated by the statistical error in the determination 
of the form factors. Although the current calculation did use a large sample of configurations, 
reducing the statistical error is certainly feasible at these values of the 
lattice spacing. This would make an improved experimental error, for example from 
BESIII, highly desirable. 

The fact the $\phi$ has a strong decay to $K\overline{K}$ and so is not 
gold-plated is an additional source of uncertainty.  Here we have estimated 
this at 3\% based on studies of the $\phi$ mass and decay constant as well 
as phenomenological arguments. 
For the $\phi$ meson, we find $M_\phi = 1.032(16)$ GeV and $f_\phi = 241(18)$ MeV, 
leading to $\Gamma(\phi \rightarrow e^+e^-) = 1.41(21) \mathrm{keV}$, in agreement 
with experiment (1.27(4) keV~\cite{pdg}). Further studies are underway of the $\phi$ with improved statistical accuracy 
and on gluon field configurations that include lighter $u/d$ quarks going down 
to physical masses~\cite{bipasha}. These should establish to higher accuracy the effect of 
the strong decay on the $\phi$ properties. 

Several elements of our calculation point the way towards future work. 
We have been able to determine the pseudoscalar form factor here which 
contributes to a lepton-mass-suppressed helicity contribution to the decay rate. 
This could be observed experimentally in the $D_s \rightarrow \phi \mu \nu$ channel 
and would give an additional handle to test weak interactions. 
We have also tested further the fact that heavy meson form factors at a given $q^2$ value
seem to be insensitive to the spectator quark mass (between light and strange masses).
Direct tests of this experimentally (for example between $D_s \rightarrow \phi$ and 
$D \rightarrow K^*$) would be interesting. 

As we have seen here and in other calculations, for example~\cite{HISQ_PRD, jpsi}, 
the HISQ action gives very small discretisation errors for $c$ quarks. This points 
the way to its use for heavier quark masses. Extrapolations to the $b$ can be 
done accurately if results at multiple lattice spacings are available, including 
very fine lattices~\cite{McNeile:2010ji, McNeile:2011ng, McNeile:2012qf}. 
The results here demonstrate that we can calculate pseudoscalar to vector meson 
transitions using HISQ valence quarks with nonperturbative normalisation of the 
vector and axial vector currents.
Working at heavier masses and finer lattices and extrapolating then gives us a new 
method for determining vector and axial vector form factors for, for example, 
$B \rightarrow D^* \ell \nu$ decays (for determination of $V_{cb}$) and $B_s \rightarrow \phi \ell^+\ell^-$ 
decays (to search for new physics). 

{\it Acknowledgements.} We are grateful to MILC for the use of their 
gauge configurations and to R. Dowdall, E. Follana, S.Playfer and P. Roudeau for useful
discussions. 
We used the Darwin Supercomputer 
as part of the DiRAC facility jointly
funded by STFC, BIS 
and the Universities of Cambridge and Glasgow. 
This work was funded by STFC.

\appendix

\section{Nonperturbative renormalisation factors for staggered bilinears}
\label{sec:zfactors}

For the staggered currents in the correlation functions needed to extract the $D_s \to \phi$ form factors and the $\phi$ decay constant, we have used a number of staggered axial vector and vector operators.
We used both a one-link vector with spin-taste $\gamma_\mu \otimes 1$ and a local vector 
 with spin-taste $\gamma_\mu \otimes \gamma_\mu$ for the $\phi$ meson 2-point correlators.
The $D_s \to \phi$ vector form factor was extracted using a local vector current ($\gamma_\mu \otimes \gamma_\mu$) for the charm to strange transition.
The axial vector form factors were extracted using both a one-link point split operator ($\gamma_5\gamma_\mu \otimes \gamma_5$) and a local axial vector operator ($\gamma_5\gamma_\mu \otimes \gamma_5\gamma_\mu$).
We have calculated renormalisation factors for each of these operators nonperturbatively.
The renormalisation factors, $Z$, on each of the ensembles used are given in Table \ref{tab:csz}.
The methods used to extract the $Z$ factors are described in the following sections.

Note that the local scalar ($1\otimes 1$) and pseudoscalar ($\gamma_5 \otimes \gamma_5$) 
operators that we use here 
are absolutely normalised through the partially conserved vector or 
axial vector current relation. This requires them to be 
multiplied by, respectively, the difference and sum of 
the lattice quark masses for the quarks appearing in the current. 
No $Z$ factor is then needed for these operators. 

\begin{table}
\begin{center}
\begin{tabular}{c|cc|ccc}
\hline
\hline
Set & $Z_{\gamma_\mu \otimes 1}^{s\bar{s}}$ & $Z_{\gamma_\mu \otimes \gamma_\mu}^{s\bar{s}}$ & $Z_{\gamma_\mu \otimes \gamma_\mu}^{c\bar{s}}$ & $Z_{\gamma_5\gamma_\mu \otimes \gamma_5}^{c\bar{s}}$ & $Z_{\gamma_5\gamma_\mu \otimes \gamma_5\gamma_\mu}^{c\bar{s}}$ \\
\hline
1 &  1.104(15) & 1.007(12) & 1.027(3) & 1.065(7) & 1.038(3) \\
\hline
2 &  1.104(15) & 1.003(9) & 1.020(10) & 1.065(5) & 1.036(4) \\
\hline
3 &  1.047(6)  & 1.009(11) & 1.009(2) & 1.017(5) & 1.020(6) \\
\hline 
\hline
\end{tabular}
\caption{The $Z$ factors on each ensemble for the staggered bilinears used. 
In columns 2 and 3 are the $Z$ factors for the $s\overline{s}$ vector currents 
used for $f_\phi$. For $Z_{\gamma_{\mu} \otimes 1}$ we use the same result, 
calculated on set 2, 
for both coarse lattices, sets 1 and 2, since we do not expect the result 
to depend significantly on the sea quark masses. In columns 4, 5 and 6, 
the $Z$ factors for 
the local vector, the 1-link axial vector and the local axial vector for the 
$c\overline{s}$ weak currents in the $D_s \to \phi \ell \nu$ calculations.}
\label{tab:csz}
\end{center}
\end{table}

\subsection{1-link vector}
The matrix element for a general pseudoscalar to pseudoscalar meson transition can be written as
\begin{align}
\langle P(p)|V_\mu|P'(p') \rangle & = f_+(q^2) \left[ p^\mu + p'^\mu - \frac{M_P^2-M_{P'}^2}{q^2}q^\mu \right] \label{eq:pstops}\\
& + f_0(q^2) \frac{M_P^2-M_{P'}^2}{q^2}q^\mu, \nonumber \end{align}
where $P$ and $P'$ are pseudoscalar mesons with momenta $p$ and $p'$ and masses $M_P$ and $M_{P'}$ respectively and $q^\mu = p'^\mu - p^\mu$.

When $q^2 = 0$, the form factors $f_+(0)$ and $f_0(0)$ are equal.
This can be used to normalise the vector current by making a 3-point function with identical mesons at the source and sink.
Eq.~\ref{eq:pstops} reduces to
\begin{equation}
\langle P(p)|V_\mu|P(p) \rangle = 2 p^\mu f_+(0).
\label{eq:pstopsrest}
\end{equation}
We normalise the $V_\mu$ operator by insisting that $f_+(0) = 1$, so we have
\begin{equation} Z \langle P(p)|V_\mu|P(p) \rangle = 2 p^\mu. \end{equation}
Here we work with a spatial vector current so this calculation must be done 
with mesons with the same non-zero momentum. 

It is particularly easy to normalise the taste-singlet vector 
operator ($\gamma_{\mu} \otimes 1$) 
in this way because a staggered propagator can be used as the spectator quark in the 3-point correlator.
The identical mesons at each end of the 3-point function must be created with operators of the same taste so we get an overall taste-singlet 3-point correlator only if we use the taste-singlet vector operator, which is a 1-link point-split current. 
To obtain the same non-zero momentum for the source and sink mesons we calculate the spectator 
quark propagator with a phased boundary condition as in Eq.~\ref{eq:twist}~\cite{etmctwist}. 
The 3-point correlator for an $\eta_s \to \eta_s$ 3-point function with this vector current inserted between strange propagators is then 
\begin{align}C_{3pt}(0,t,T) & = \sum_{x,y,z}(-1)^{y_\mu^<} \varepsilon(y)\varepsilon(z) \label{eq:pstopscorr} \\
 & \times \mathrm{Tr}\left[g_s^{\theta}(x,z) g_s(z,y) g^{\dag}_s(x,y\pm \hat{\mu})\right]. \nonumber \end{align}
As before, the sites $x$, $y$ and $z$ are at times $0$, $t$ and $T$ respectively and we sum over lattice sites on the same timeslice.

To obtain the $Z$ factors the 3-point correlators are fitted, along with 
the appropriate 2-point correlators, according to the 
fit form given in Eq.~\ref{eq:3ptfit}.
For the $\eta_s$ we use the Goldstone pseudoscalar operator so that the 
2-point correlator is simply the modulus squared of the strange quark 
propagator. 
Since the 3-point function is symmetric in this case, with identical 
mesons at 0 and $T$, we can impose on the fit that $V^{nn}$ 
and $V^{oo}$ are symmetric matrices and $V^{on} = V^{no}$.

The $Z$ factors obtained for the 1-link taste-singlet vector current  
are given in Table~\ref{tab:csz}~\cite{jklatt11} and are labelled 
as $Z_{\gamma_\mu \otimes 1}^{s\bar{s}}$.
Note that the $Z$ factors in this case are significantly different from 1. 
The $Z$ factor was shown in \cite{jklatt11} to be independent of the 
spatial momentum used for the spectator quark and of whether the spectator 
was a charm or strange quark (i.e. comparing $\eta_s \rightarrow \eta_s$ with 
$D_s \rightarrow D_s$).  
We also found that the $Z$ factor did not change significantly between a $c\overline{c}$ and 
$s\overline{s}$ current. 

\subsection{Local vector}
We use a local vector current for $f_\phi$ with equal mass (both strange) quarks. 
We also use the same operator with unequal quark masses for the charm to strange transition in $D_s \to \phi$. 
The $Z$ factors for these two cases are calculated with different methods.
In both cases, it is simplest to normalise the temporal vector; for the relativistic 
HISQ action the renormalisation factors for the spatial and temporal 
components will differ only by discretisation effects which vanish in the 
continuum limit. For a temporal vector current renormalisation we can work 
with mesons at rest. 

\subsubsection{Equal quark mass case}
\begin{table}
\begin{center}
\begin{tabular}{cccc}
\hline
\hline
Set &  $am_h$ & $T$ values & $Z$ \\
\hline
1 &  2.0 & 15,16,20,21 & 1.007(12) \\
\hline
2 &  2.0 & 15,16,20,21 & 1.003(9) \\
 & 2.8 & 15,16,20,21 & 0.996(13) \\
\hline
3 & 1.5 & 24,25,30,31 & 1.009(11) \\
\hline 
\hline
\end{tabular}
\caption{Further details for the calculation of the $Z$ factors on each ensemble for the local $\bar{s} (\gamma_\mu \otimes \gamma_\mu) s$ operator. 4 time sources were used per configuration. The NRQCD masses used are given in column 2. For ensemble 2, two masses were used for the NRQCD spectator quark. Column 3 gives the values of $T$ used for the 3-point correlators and column 4 gives the $Z$ factor.}
\label{tab:localstrange}
\end{center}
\end{table}

As the local vector operator is not a taste-singlet, it cannot simply be inserted into a pseudoscalar to pseudoscalar symmetric 3-point function where staggered quarks are used for each of the propagators.
However, it can be normalised using a 3-point function where the spectator quark retains 4 spin components.
Here it is convenient to use NRQCD for the spectator quark, as we did for the 
local charm-charm vector operator in~\cite{jpsi}.
The staggered-staggered current renormalisation factor should not depend (up to discretisation effects) on the details of the spectator quark, so there is no need for the NRQCD quark mass to correspond to a physical quark.

To combine a staggered propagator with one carrying spin indices, we convert the 
staggered propagator to a 4-spin naive propagator using the products of 
$\gamma$ matrices (here denoted $\Omega$) which 
diagonalise the naive quark action in spin space~\cite{HISQ_PRD}.
A staggered-NRQCD 2-point correlator is given by
\begin{eqnarray} C_{2pt}(0,t) &=& \\ &&\sum_{x,y} \mbox{Tr} \{ G_{\mathrm{{NRQCD}}}(x,y) \Omega(y) g^\dag(x,y) \Omega^\dag(x)  \}, \nonumber\end{eqnarray}
where sites $x$ and $y$ are at times $0$ and $t$, the sum is over timeslices and the trace is over both spin and colour.
The staggered propagator $g(x,y)$ contains no spin dependence 
and the $\Omega$ matrices contain no colour,
so that the traces in the correlator can be separated:
\begin{eqnarray} && C_{2pt}(0,t) = \\ &&\sum_{x,y} \mbox{Tr}_c \{ \mbox{Tr}_s [ \Omega^\dag(x) G_{\mathrm{NRQCD}}(x,y) \Omega(y) ] g^\dag(x,y)   \}. \nonumber\end{eqnarray}
This shows that we can take the spin trace after multiplying the NRQCD propagator by the $\Omega$ matrices and we can do this before combining it with the staggered propagator.

We can extend this to 3-point functions and take the spin trace of the NRQCD propagator and $\gamma$ matrices in the middle of the calculation.
The 3-point correlation function is 
\begin{align} & C_{3pt}(0,t,T)  = \sum_{x,y,z} \varepsilon(z)(-1)^{y_t} \label{nrqcdstrange} \\ & \times \mathrm{Tr_c}\left\{ \mathrm{Tr_s}\left[\gamma_t\Omega^\dagger(x)G_{\mathrm{NRQCD}}(x,z)\Omega(z)\right] g(z,y) g^\dagger(x,y) \right\}, \nonumber  \end{align}
where sites $x$, $y$ and $z$ are at times $0$, $t$ and $T$.
The NRQCD propagator is the spectator (propagator 1 in Fig. \ref{fig:3ptdiag}).
We use  $\sum_x\varepsilon(z)\mathrm{Tr_s}\left[\gamma_t\Omega^\dagger(x)G_{\mathrm{NRQCD}}(x,z)\Omega(z) \right]$ as the source for the inversion of extended propagator 2.

We calculate the NRQCD-HISQ 2-point and 3-point functions for different $T$ values.
Details of the parameters used and results are given in Table \ref{tab:localstrange}.
Up to discretisation effects, $Z$ should not depend on the mass of the spectator quark.
On ensemble 2, the $Z$ factor was calculated using two values of the heavy NRQCD quark mass, $am_h$, and the results agree within the statistical errors.
The $Z$ factors used to normalise the current for $f_\phi$ are the ones obtained with $am_h = 2.0$ for the coarse ensembles 1 and 2 and $am_h = 1.5$ for the fine ensemble 3.
These NRQCD masses correspond to approximately the same physical quark mass.

We summarise our results for $Z_{\gamma_\mu \otimes \gamma_\mu}^{s\bar{s}}$ in Table \ref{tab:csz}. Values are close to 1 for this vector operator. 

\subsubsection{Unequal quark mass case}
The local temporal $c\overline{s}$ vector can be normalised 
using a local non-Goldstone $D_s$ 
(made with a $\gamma_t\gamma_5 \otimes \gamma_t \gamma_5$ operator) 
in a $D_s \to \eta_s$ 3-point correlator 
with both the $D_s$ and $\eta_s$ at rest. 
From Eq.~\ref{eq:pstops} we see that the matrix element is 
then given by $f_0(q_{max}^2)(M_P + M_{P^{\prime}})$, up to 
a $Z$ factor for the vector current. 
By comparing this to the result from the absolutely 
normalised local scalar current between the Goldstone $D_s$ 
and $\eta_s$ we can extract $Z$~\cite{jonnadtok, Na:2010uf}. 
The scalar current matrix element is given by: 
\begin{equation}
\langle P(p)|S|P'(p') \rangle  = f_0(q^2) \frac{M_{P'}^2-M_{P}^2}{m_{01}-m_{02}}
\label{eq:sff}
\end{equation}
where $P'$ is a $D_s$ meson, $P$, an $\eta_s$, $m_{01}$ is the lattice 
charm quark mass and $m_{02}$, the lattice strange quark mass. 

The difference in mass between the Goldstone and non-Goldstone $D_s$ is a small lattice artefact which will mean that $q^2_{max}$ is not quite the same 
in the two cases. 
These two masses appear in Table \ref{tab:dsphidata}; we use a Goldstone $D_s$ in our extraction of the axial vector form factors and the non-Goldstone $D_s$ is used when calculating the vector form factor.
We see that the difference between them is very small even on the coarse 
lattices, and is clearly vanishing rapidly as the lattice spacing goes 
to zero. 

The renormalisation factors that we obtain from this method are given in 
Table \ref{tab:csz} as $Z_{\gamma_\mu \otimes \gamma_\mu}^{c\bar{s}}$.
Similarly to the equal mass case, they have values close to 1. 

\subsection{1-link axial vector}
\label{sec:1linkamu}
The 1-link axial vector operator ($\gamma_{\mu} \gamma_5 \otimes \gamma_5$) 
that we use includes a point-splitting 
in the same spatial direction as the polarization of the axial vector.
It therefore has the same taste as local pseudoscalar operator and 
the partially conserved axial current, and we can use this to 
normalise it. 
For the HISQ action, the partially conserved axial current 
relation gives
\begin{equation}p_\mu \langle 0 |A_\mu| P_0 \rangle = (m_{01}+m_{02}) \langle 0|\gamma_5|P_0 \rangle \label{stagpcac}\end{equation}
for a pseudoscalar meson $P_0$ with valence quarks of lattice 
quark masses $m_{01}$ and $m_{02}$. 

\begin{figure}
\centering
\includegraphics[angle=-90,width=0.45\textwidth]{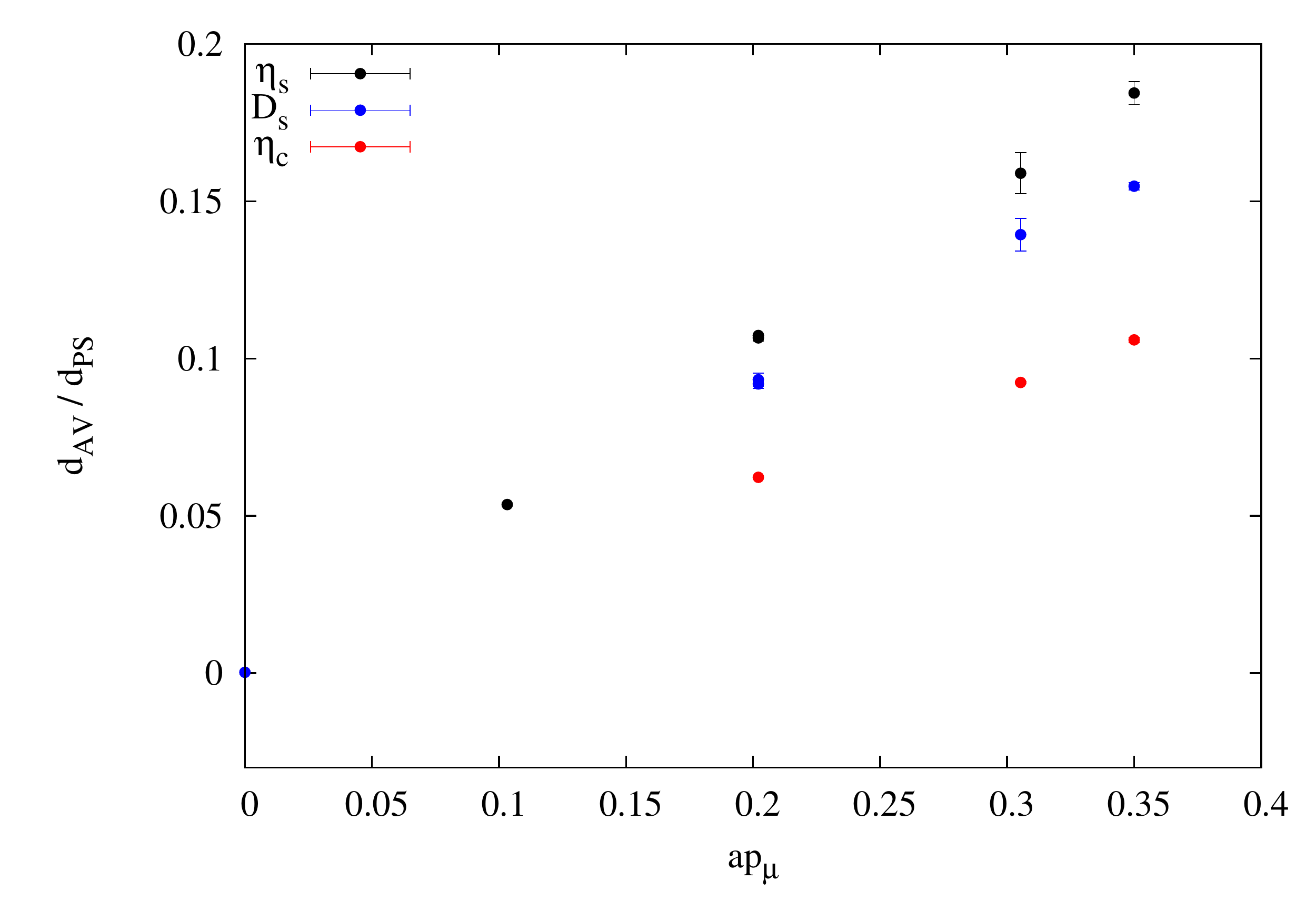}
\caption{The ratio of axial vector and pseudoscalar amplitudes plotted 
against meson momentum in lattice units 
for pseudoscalar mesons made from $c$ and $s$
valence quarks. The results come from coarse lattices, set 2. 
This ratio is proportional to the momentum and 
the renormalisation factor for the axial vector operator can be 
extracted from the gradient of the line, given 
by $\frac{m_{01}+m_{02}}{M_{P_0}^2}\frac{1}{Z}$.}
\label{fig:1linkamu}
\end{figure}

We can then normalise the 1-link axial vector operator using a 
correlation function where $P_0$ is created using a local 
pseudoscalar and destroyed with a 1-link axial vector.
The correlator is
\begin{equation}C_{PS\to AV} (0,t) = i\sum_{x,y} (-1)^{y_\mu^<} \times \mathrm{Tr}\left[ g_1(x,y) g_2^\dag(x,y\pm\hat{\mu}) \right] \label{our1linkoperator}\end{equation}
where $x$ is at time $0$ and $y$ at $t$.
The point-splitting at $y$ is implemented by averaging over links in the forward and backward directions. 
We fit simultaneously with the pseudoscalar 2-point correlator where $P_0$ is created and destroyed by the same local pseudoscalar operator, given by
\begin{equation}C_{PS\to PS}(0,t) = \sum_{x,y} \mathrm{Tr}\left[ g_1(x,y) g_2^\dag(x,y) \right].\end{equation}

The form of the fit used is given for the 2-point function in 
Eq.~\ref{eq:3ptfit}. Since we have different operators at the 
source and sink for the $C_{PS \to AV}$ correlators, the amplitudes 
for each state in that case are the product of a source operator amplitude  
and a sink operator amplitude. 
Thus, for the ground-state, the amplitude in $C_{PS\to PS}$ is 
$d_{PS}^2$ and for $C_{PS\to AV}$ it is $d_{PS}d_{AV}$. 

Including the $Z$ factor for the operator, the axial vector fit 
amplitude $d_{AV}$ is then related to the pseudoscalar 
amplitude $d_{PS}$ through Eq.~\ref{stagpcac} by
\begin{equation} Z d_{AV} = \frac{m_{01}+m_{02}}{M_{P_0}^2}p_\mu d_{PS}. \label{1linkz}\end{equation}
Because the $\mu$ direction is spatial here, we need to include momentum in the meson to normalise the axial vector operator in this way.

The ratio $d_{AV}/d_{PS}$ is plotted against the meson momentum for pseudoscalar mesons containing charm and strange quarks for coarse set 2 
in Figure \ref{fig:1linkamu}.
From Eq.~\ref{1linkz}, the ratio should be proportional to $p_\mu$, where 
the $\mu$ is the direction of the axial vector, and we see that 
the results indeed do give straight lines through the origin.
The $Z$ factor can then be extracted from the gradient, 
which is $(m_{01}+m_{02})/(M_{P_0}^2Z)$.

If the quark masses are unequal and we use the 1-link operator in Eq.~\ref{our1linkoperator}, then we find a dependence on which quark propagator carries 
the meson's momentum.
For the $c\overline{s}$ current that we use, the operator is normalised with the $s$ quark carrying the momentum. 
This is the same situation as appears in the $D_s \to \phi$ 3-point functions.
The effects of including momentum in point-split operators are discussed 
further in Appendix~\ref{ptsplitmomentum}. 
We find that the dependence on which quark carries the momentum is a lattice spacing artefact.

The values we obtain for $Z_{\gamma_5\gamma_\mu \otimes \gamma_5}^{c\bar{s}}$ 
for each ensemble are given in Table~\ref{tab:csz}.

\subsection{Local axial vector}
The temporal component of the local axial vector ($\gamma_5\gamma_t \otimes \gamma_5 \gamma_t$) can be normalised by comparing amplitudes 
for Goldstone and local non-Goldstone pseudoscalar meson correlators.
To normalise the $c\overline{s}$ current which appears in the charm 
to strange decay, we simply demand that the matrix element for 
the temporal axial current be the same for Goldstone and 
non-Goldstone $D_s$ mesons.

From fits to the separate Goldstone and non-Goldstone correlators 
we obtain amplitudes for the ground-state of $d_{PS}$ and 
$d_{LTAV}$ respectively. And then, from Eq.~\ref{1linkz} but 
for the temporal case at zero momentum, we have normalisation 
condition
\begin{equation} Z d_{LTAV} = \frac{(m_{0c}+m_{0s})}{M_{D_s}}d_{PS}. \end{equation} 
As before, a small discretisation effect arises from the fact that 
the masses of the Goldstone and non-Goldstone $D_s$ are not exactly 
the same at non-zero lattice spacing. 
The renormalisation factor, $Z_{\gamma_5\gamma_\mu \otimes \gamma_5\gamma_\mu}^{c\bar{s}}$, is easily extracted and given in Table \ref{tab:csz}.

\section{Point-split operators with momentum}
\label{ptsplitmomentum}
Here we consider an issue with momentum and point split operators, which arises for the pseudoscalar to axial vector correlators we consider in Appendix~\ref{sec:1linkamu}.
For quark propagators carrying momentum, we use boundary conditions that 
incorporate a phase as in Eq.~\ref{eq:twist}~\cite{firsttwist, etmctwist}.
A propagator $g_1^{\theta_a}(x,y)$ for a quark with mass $m_1$ carrying momentum $p^a$ is related to propagator calculated without twisted boundary conditions by
\begin{equation} g_1^{\theta_a}(x,y) = g_1(x,y) e^{-i \theta_a(x-y)}	\end{equation} 
where $p_a = \theta_a/L_s$ for a lattice of spatial size $L_s$.

\begin{figure}
\centering
\includegraphics[angle=-90,width=0.45\textwidth]{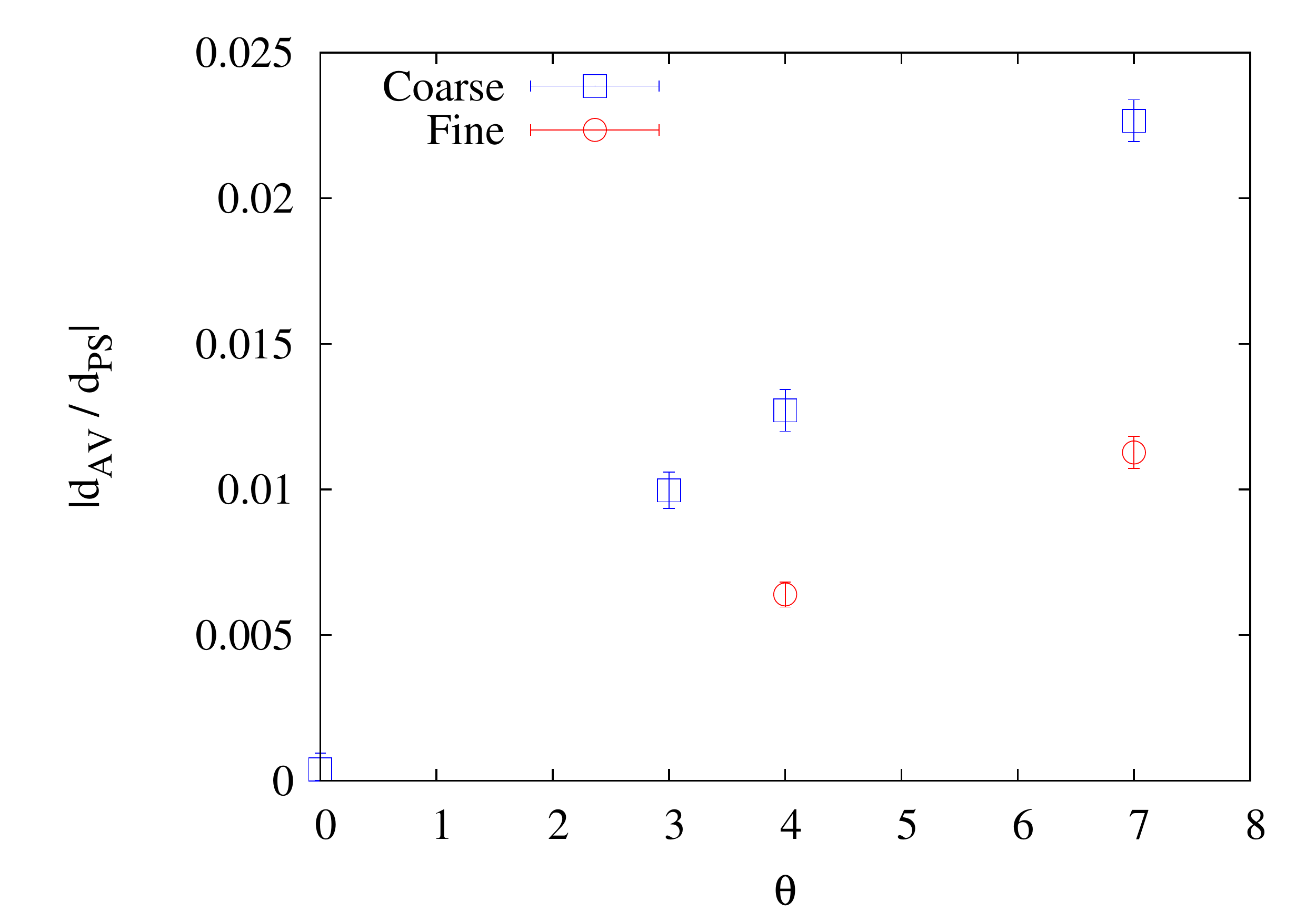}
\caption{The ratio of the axial vector to pseudoscalar amplitudes in the case where the charm and strange propagators carry the same phase at the boundary, $\theta$. 
In this case, the total meson momentum is zero and we expect this ratio to be zero. The ratio is plotted against the momentum carried (in opposite directions) by each propagator. The blue squares are for the coarse Set 2 and the red circles for fine Set 3. These gauge configurations have approximately the same physical size, so the same $\theta$ corresponds to the same physical momentum on each.}
\label{fig:thetafinecoarse}
\end{figure}

\begin{figure}
\centering
\includegraphics[angle=-90,width=0.45\textwidth]{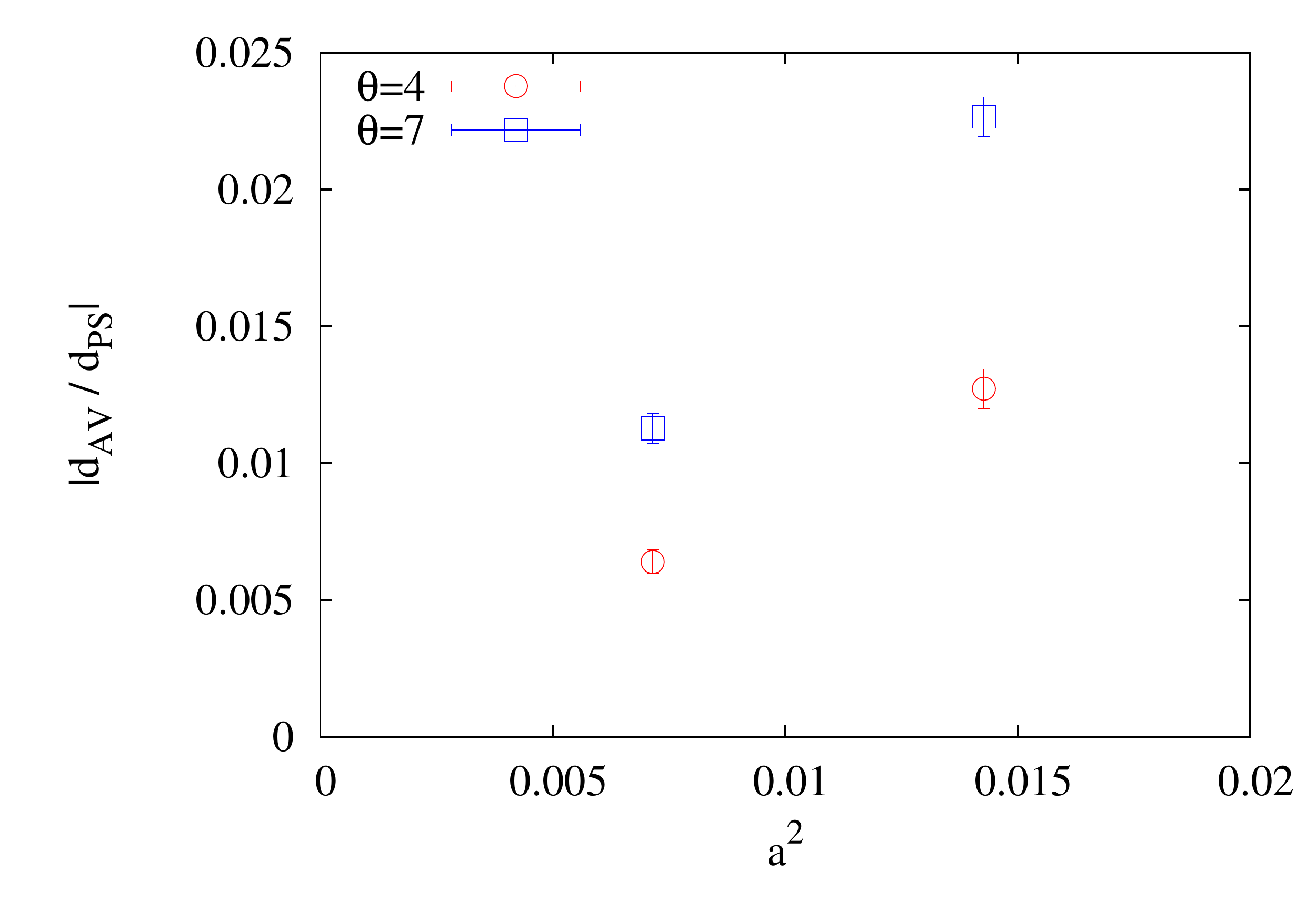}
\caption{The ratio of the axial vector to pseudoscalar amplitudes in the case where the charm and strange propagators carry the same twist. In this case, the total meson momentum is zero and we expect this ratio to be zero. The ratio is plotted against $a^2$. Sets 2 and 3 have different lattice spacing, but the same physical size so the same twist $\theta$ on each corresponds to the same physical momentum. The red and blue data points are for the cases in which both the strange and charm propagators carry $\theta = 4,7$.}
\label{fig:thetafinecoarsea2}
\end{figure}

\begin{figure}
\centering
\includegraphics[angle=-90,width=0.45\textwidth]{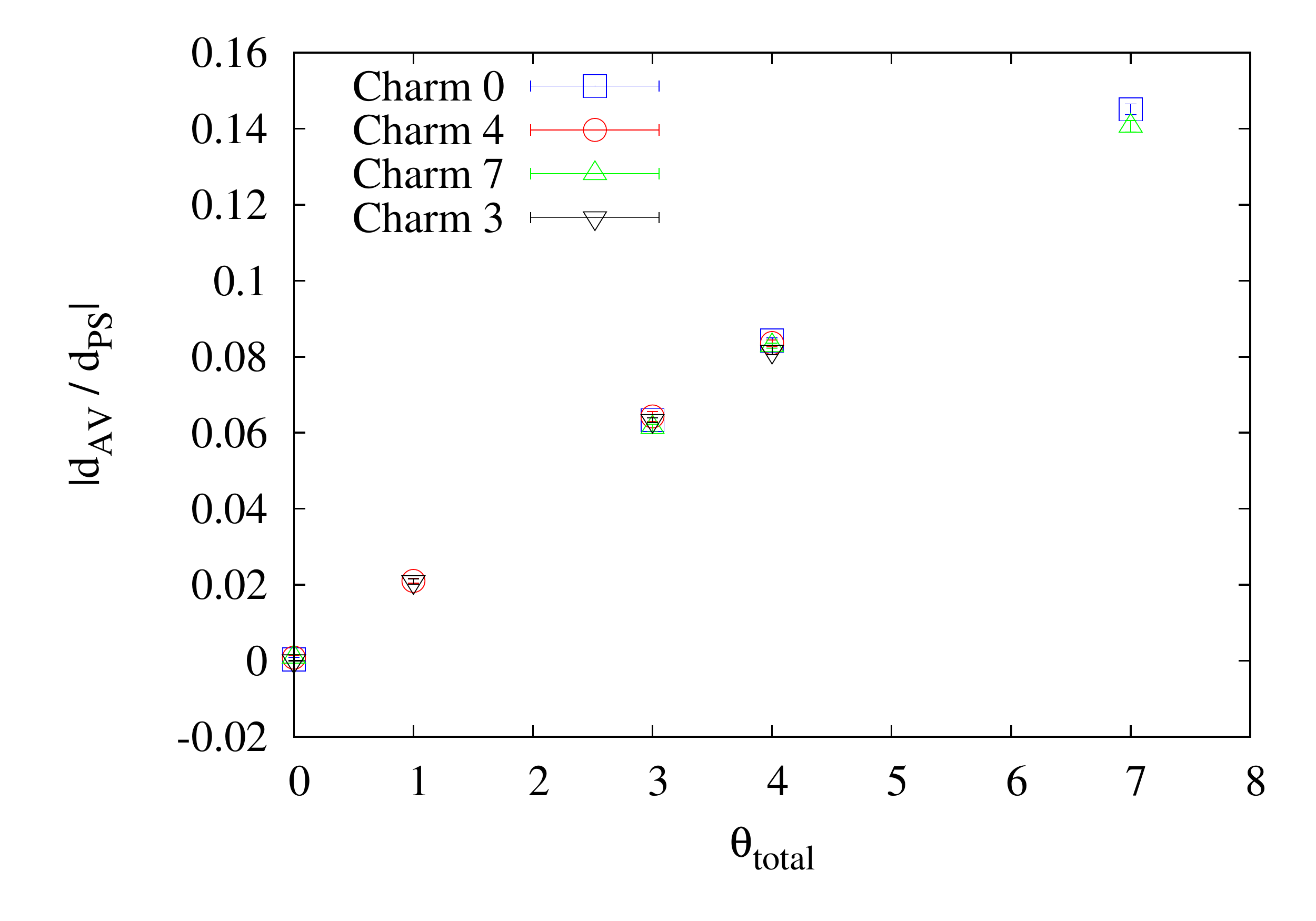}
\caption{The ratio of axial vector to pseudoscalar fit amplitudes plotted against $D_s$ 
total momentum (given by the difference of phases carried 
by the $c$ and $s$ quarks), where the 1-link axial vector 
operator is given by Eq.~\ref{alt1link}. 
The different symbols indicate the momentum (phase at the boundary) 
carried by the charm quark. Results are for the coarse lattices, set 2. 
Now the amplitude ratio agrees for points where the meson total momentum is same.}
\label{fig:avcorrected}
\end{figure}

For the 1-link spatial axial vector current, we implement a symmetric point-splitting at the sink in, say, the $\mu$ direction and write the correlator as
\begin{equation} \frac{i}{2} g^{\theta_1\dag}_1(x,y) \left\{ g_2^{\theta_2}(x,y+\hat{\mu}) + g_2^{\theta_2}(x,y-\hat{\mu}) \right\} \label{normal1link} \end{equation}
omitting the sum over timeslices, staggered phase and colour trace.
Note that the staggered phase in Eq.~\ref{our1linkoperator} does not contain $(-1)^{y_\mu}$ which means that the phase factor is the same at $y$ and $y\pm\hat{\mu}$. 

The meson made of quarks 1 and 2 has total momentum $\theta_{total} = \theta_{2} - \theta_{1}$ 
so the twisted boundary condition only needs to be applied in one of the inversions.
Also different combinations of $\theta_1$ and $\theta_2$ should give the 
same result for the same $\theta_{total}$. 
When the two quarks have equal mass it makes no difference, but in the unequal 
mass case the simple operator in Eq.~\ref{normal1link} shows results that depend 
on which quark carries which momentum. 

The simplest case to study is that in which the total momentum given by 
$\theta$ is zero. Then the amplitude of the spatial axial vector current 
between the vacuum and a pseudoscalar meson 
(i.e. $d_{AV}$ in Appendix~\ref{sec:1linkamu}) should be zero.  
Putting momentum $\theta_2=\theta_1=\theta$ and writing Equation \ref{normal1link} in terms of propagators with no twist gives
\begin{equation} \frac{i}{2} g^\dag_1(x,y) \left\{ g_2(x,y+\hat{\mu})e^{i\theta} + g_2(x,y-\hat{\mu})e^{-i\theta} \right\}. \label{thetaq}\end{equation} 
This should give zero for all values of $\theta$ when summed over $y$. In the equal mass 
case this is true because the two pieces are complex conjugates of each other 
so the real part of the correlator is zero (configuration by configuration). 

In the unequal mass case the amplitude is not zero and depends on $\theta$. 
In Figure \ref{fig:thetafinecoarse}, we plot the ratio for $D_s$ mesons 
created this way on both fine and coarse gauge configurations.
The data is from Sets 2 and 3 with approximately the same lattice length in physical 
units ($aL_s$) so the same $\theta$ on each corresponds to the same physical momentum.
We see that the difference of the ratio from zero depends linearly on the 
momentum carried by each quark, denoted by $\theta$.
We also see, however, that the difference is less on the fine lattices than on the coarse.
A similar situation holds for non-zero meson momentum in that the 
amplitude ratio shows a spread which depends on how that momentum 
is made up from the quark momenta. 

In Figure~\ref{fig:thetafinecoarsea2}, we plot the same zero momentum 
meson data as in Figure~\ref{fig:thetafinecoarse}, but now against $a^2$.
For each value of $\theta$, we see that the discrepancy depends on $a^2$.
This demonstrates that the ambiguity in which quark 
carries the meson's momentum is a discretisation error.
In our results (as discussed in Appendix~\ref{sec:1linkamu}) we have determined 
$Z$ using twisted boundary conditions for the $s$ quark. The results above 
show that the same results would be obtained in the continuum limit if instead 
we had used twisted boundary conditions for the $c$ quarks.  

The difference can be avoided by using a definition of the 
point-split operator that is less affected by they way in which 
the momentum is split between the propagators, 
and is more symmetric. 
One example would be to combine propagators $g^{\theta_1}_1(x,y)$ 
carrying momentum $\theta_1$ and $g^{\theta_2}_2(x,y)$ carrying $\theta_2$ in 
the following way:
\begin{align} & \frac{i}{4} \biggl\{ g^{\theta_1\dag}_1(x,y) g_2^{\theta_2}(x,y+\hat{\mu}) e^{-i\theta_2} \nonumber \\ 
& + g^{\theta_1\dag}_1(x,y+\hat{\mu}) g_2^{\theta_2}(x,y) e^{i\theta_1} \nonumber \\
& + g^{\theta_1\dag}_1(x,y) g_2^{\theta_2}(x,y-\hat{\mu}) e^{i\theta_2} \nonumber \\ 
& + g^{\theta_1\dag}_1(x,y-\hat{\mu}) g_2^{\theta_2}(x,y) e^{-i\theta_1} \biggr\} \label{alt1link} \end{align}
which has meson momentum $\theta_{total}=\theta_2-\theta_1$.
If the meson is to carry momentum $\theta$ using only a phase in the quark propagator $g_2(x,y)$, then we have $\theta_1=0, \theta_2 = \theta$ and if the momentum is carried only by $g_1(x,y)$ then $\theta_1 = -\theta, \theta_2 =0$.
In either case, writing Eq.~\ref{alt1link} in terms of propagators calculated without twisted boundary conditions gives
\begin{align} & \frac{ie^{-i\theta(x-y)}}{4} \biggl\{ g^\dag_1(x,y) g_2(x,y+\hat{\mu}) + g^\dag_1(x,y+\hat{\mu}) g_2(x,y) \nonumber \\
 & + g^\dag_1(x,y) g_2(x,y-\hat{\mu}) + g^\dag_1(x,y-\hat{\mu}) g_2(x,y) \biggr\}. \label{alttheta} \end{align}
This is also true for all $\theta_1$ and $\theta_2$ that satisfy $\theta_2 - \theta_1 = \theta$.

Using this form of the 1-link axial vector operator gives the axial vector to 
pseudoscalar amplitude ratios for the $D_s$ shown in Figure~\ref{fig:avcorrected}.
Now we see that the amplitude is zero at zero meson momentum and 
we have good agreement between amplitude ratios for a given meson 
momentum that correspond to a different distribution of momentum 
between $s$ and $c$ quarks. 

\bibliography{dsphi}

\begin{thebibliography}{31}
\expandafter\ifx\csname natexlab\endcsname\relax\def\natexlab#1{#1}\fi
\expandafter\ifx\csname bibnamefont\endcsname\relax
  \def\bibnamefont#1{#1}\fi
\expandafter\ifx\csname bibfnamefont\endcsname\relax
  \def\bibfnamefont#1{#1}\fi
\expandafter\ifx\csname citenamefont\endcsname\relax
  \def\citenamefont#1{#1}\fi
\expandafter\ifx\csname url\endcsname\relax
  \def\url#1{\texttt{#1}}\fi
\expandafter\ifx\csname urlprefix\endcsname\relax\def\urlprefix{URL }\fi
\providecommand{\bibinfo}[2]{#2}
\providecommand{\eprint}[2][]{\url{#2}}

\bibitem[{\citenamefont{Koponen et~al.}(2013)\citenamefont{Koponen, Davies,
  Donald, Follana, Lepage et~al.}}]{jonnadtok}
\bibinfo{author}{\bibfnamefont{J.}~\bibnamefont{Koponen}},
  \bibinfo{author}{\bibfnamefont{C.}~\bibnamefont{Davies}},
  \bibinfo{author}{\bibfnamefont{G.}~\bibnamefont{Donald}},
  \bibinfo{author}{\bibfnamefont{E.}~\bibnamefont{Follana}},
  \bibinfo{author}{\bibfnamefont{G.}~\bibnamefont{Lepage}},
  \bibnamefont{et~al.} (\bibinfo{collaboration}{HPQCD Collaboration})
  (\bibinfo{year}{2013}), \eprint{1305.1462}.

\bibitem[{\citenamefont{Follana et~al.}(2007)}]{HISQ_PRD}
\bibinfo{author}{\bibfnamefont{E.}~\bibnamefont{Follana}} \bibnamefont{et~al.}
  (\bibinfo{collaboration}{HPQCD and UKQCD Collaborations}),
  \bibinfo{journal}{Phys.Rev.} \textbf{\bibinfo{volume}{D75}},
  \bibinfo{pages}{054502} (\bibinfo{year}{2007}), \eprint{hep-lat/0610092}.

\bibitem[{\citenamefont{Richman and Burchat}(1995)}]{richman}
\bibinfo{author}{\bibfnamefont{J.~D.} \bibnamefont{Richman}} \bibnamefont{and}
  \bibinfo{author}{\bibfnamefont{P.~R.} \bibnamefont{Burchat}},
  \bibinfo{journal}{Rev.Mod.Phys.} \textbf{\bibinfo{volume}{67}},
  \bibinfo{pages}{893} (\bibinfo{year}{1995}), \eprint{hep-ph/9508250}.

\bibitem[{\citenamefont{Aubert et~al.}(2008)}]{babar}
\bibinfo{author}{\bibfnamefont{B.}~\bibnamefont{Aubert}} \bibnamefont{et~al.}
  (\bibinfo{collaboration}{BaBar Collaboration}), \bibinfo{journal}{Phys.Rev.}
  \textbf{\bibinfo{volume}{D78}}, \bibinfo{pages}{051101}
  (\bibinfo{year}{2008}), \eprint{0807.1599}.

\bibitem[{\citenamefont{Korner and Schuler}(1990)}]{Korner:1989qb}
\bibinfo{author}{\bibfnamefont{J.}~\bibnamefont{Korner}} \bibnamefont{and}
  \bibinfo{author}{\bibfnamefont{G.}~\bibnamefont{Schuler}},
  \bibinfo{journal}{Z.Phys.} \textbf{\bibinfo{volume}{C46}},
  \bibinfo{pages}{93} (\bibinfo{year}{1990}).

\bibitem[{\citenamefont{Briere et~al.}(2010)}]{cleodkstar}
\bibinfo{author}{\bibfnamefont{R.}~\bibnamefont{Briere}} \bibnamefont{et~al.}
  (\bibinfo{collaboration}{CLEO Collaboration}), \bibinfo{journal}{Phys.Rev.}
  \textbf{\bibinfo{volume}{D81}}, \bibinfo{pages}{112001}
  (\bibinfo{year}{2010}), \eprint{1004.1954}.

\bibitem[{\citenamefont{Follana et~al.}(2008)\citenamefont{Follana, Davies,
  Lepage, and Shigemitsu}}]{HISQ_PRL}
\bibinfo{author}{\bibfnamefont{E.}~\bibnamefont{Follana}},
  \bibinfo{author}{\bibfnamefont{C.}~\bibnamefont{Davies}},
  \bibinfo{author}{\bibfnamefont{G.}~\bibnamefont{Lepage}}, \bibnamefont{and}
  \bibinfo{author}{\bibfnamefont{J.}~\bibnamefont{Shigemitsu}}
  (\bibinfo{collaboration}{HPQCD and UKQCD Collaborations}),
  \bibinfo{journal}{Phys.Rev.Lett.} \textbf{\bibinfo{volume}{100}},
  \bibinfo{pages}{062002} (\bibinfo{year}{2008}), \eprint{0706.1726}.

\bibitem[{\citenamefont{Davies et~al.}(2010{\natexlab{a}})\citenamefont{Davies,
  McNeile, Follana, Lepage, Na et~al.}}]{Dsdecayconst}
\bibinfo{author}{\bibfnamefont{C.}~\bibnamefont{Davies}},
  \bibinfo{author}{\bibfnamefont{C.}~\bibnamefont{McNeile}},
  \bibinfo{author}{\bibfnamefont{E.}~\bibnamefont{Follana}},
  \bibinfo{author}{\bibfnamefont{G.}~\bibnamefont{Lepage}},
  \bibinfo{author}{\bibfnamefont{H.}~\bibnamefont{Na}}, \bibnamefont{et~al.}
  (\bibinfo{collaboration}{HPQCD Collaboration}), \bibinfo{journal}{Phys.Rev.}
  \textbf{\bibinfo{volume}{D82}}, \bibinfo{pages}{114504}
  (\bibinfo{year}{2010}{\natexlab{a}}), \eprint{1008.4018}.

\bibitem[{\citenamefont{Donald et~al.}(2012)\citenamefont{Donald, Davies,
  Dowdall, Follana, Hornbostel et~al.}}]{jpsi}
\bibinfo{author}{\bibfnamefont{G.}~\bibnamefont{Donald}},
  \bibinfo{author}{\bibfnamefont{C.}~\bibnamefont{Davies}},
  \bibinfo{author}{\bibfnamefont{R.}~\bibnamefont{Dowdall}},
  \bibinfo{author}{\bibfnamefont{E.}~\bibnamefont{Follana}},
  \bibinfo{author}{\bibfnamefont{K.}~\bibnamefont{Hornbostel}},
  \bibnamefont{et~al.} (\bibinfo{collaboration}{HPQCD Collaboration}),
  \bibinfo{journal}{Phys.Rev.} \textbf{\bibinfo{volume}{D86}},
  \bibinfo{pages}{094501} (\bibinfo{year}{2012}), \eprint{1208.2855}.

\bibitem[{\citenamefont{Bazavov et~al.}(2010)\citenamefont{Bazavov, Toussaint,
  Bernard, Laiho, DeTar et~al.}}]{MILCconfigs}
\bibinfo{author}{\bibfnamefont{A.}~\bibnamefont{Bazavov}},
  \bibinfo{author}{\bibfnamefont{D.}~\bibnamefont{Toussaint}},
  \bibinfo{author}{\bibfnamefont{C.}~\bibnamefont{Bernard}},
  \bibinfo{author}{\bibfnamefont{J.}~\bibnamefont{Laiho}},
  \bibinfo{author}{\bibfnamefont{C.}~\bibnamefont{DeTar}},
  \bibnamefont{et~al.}, \bibinfo{journal}{Rev.Mod.Phys.}
  \textbf{\bibinfo{volume}{82}}, \bibinfo{pages}{1349} (\bibinfo{year}{2010}),
  \eprint{0903.3598}.

\bibitem[{\citenamefont{Davies et~al.}(2010{\natexlab{b}})\citenamefont{Davies,
  Follana, Kendall, Lepage, and McNeile}}]{Davies:2009tsa}
\bibinfo{author}{\bibfnamefont{C.}~\bibnamefont{Davies}},
  \bibinfo{author}{\bibfnamefont{E.}~\bibnamefont{Follana}},
  \bibinfo{author}{\bibfnamefont{I.}~\bibnamefont{Kendall}},
  \bibinfo{author}{\bibfnamefont{G.}~\bibnamefont{Lepage}}, \bibnamefont{and}
  \bibinfo{author}{\bibfnamefont{C.}~\bibnamefont{McNeile}}
  (\bibinfo{collaboration}{HPQCD Collaboration}), \bibinfo{journal}{Phys.Rev.}
  \textbf{\bibinfo{volume}{D81}}, \bibinfo{pages}{034506}
  (\bibinfo{year}{2010}{\natexlab{b}}), \eprint{0910.1229}.

\bibitem[{\citenamefont{Gregory et~al.}(2011)}]{Gregory:2010gm}
\bibinfo{author}{\bibfnamefont{E.~B.} \bibnamefont{Gregory}}
  \bibnamefont{et~al.} (\bibinfo{collaboration}{HPQCD collaboration}),
  \bibinfo{journal}{Phys. Rev.} \textbf{\bibinfo{volume}{D83}},
  \bibinfo{pages}{014506} (\bibinfo{year}{2011}), \eprint{1010.3848}.

\bibitem[{\citenamefont{Naik}(1989)}]{naik}
\bibinfo{author}{\bibfnamefont{S.}~\bibnamefont{Naik}},
  \bibinfo{journal}{Nucl.Phys.} \textbf{\bibinfo{volume}{B316}},
  \bibinfo{pages}{238} (\bibinfo{year}{1989}).

\bibitem[{\citenamefont{Bazavov et~al.}(2012)}]{milc-hisq}
\bibinfo{author}{\bibfnamefont{A.}~\bibnamefont{Bazavov}} \bibnamefont{et~al.}
  (\bibinfo{collaboration}{MILC Collaboration}) (\bibinfo{year}{2012}),
  \eprint{1212.4768}.

\bibitem[{\citenamefont{McNeile et~al.}(2002)\citenamefont{McNeile, Michael,
  and Sharkey}}]{McNeile:2001cr}
\bibinfo{author}{\bibfnamefont{C.}~\bibnamefont{McNeile}},
  \bibinfo{author}{\bibfnamefont{C.}~\bibnamefont{Michael}}, \bibnamefont{and}
  \bibinfo{author}{\bibfnamefont{K.}~\bibnamefont{Sharkey}}
  (\bibinfo{collaboration}{UKQCD Collaboration}), \bibinfo{journal}{Phys.Rev.}
  \textbf{\bibinfo{volume}{D65}}, \bibinfo{pages}{014508}
  (\bibinfo{year}{2002}), \eprint{hep-lat/0107003}.

\bibitem[{\citenamefont{de~Divitiis et~al.}(2004)\citenamefont{de~Divitiis,
  Petronzio, and Tantalo}}]{firsttwist}
\bibinfo{author}{\bibfnamefont{G.}~\bibnamefont{de~Divitiis}},
  \bibinfo{author}{\bibfnamefont{R.}~\bibnamefont{Petronzio}},
  \bibnamefont{and} \bibinfo{author}{\bibfnamefont{N.}~\bibnamefont{Tantalo}},
  \bibinfo{journal}{Phys.Lett.} \textbf{\bibinfo{volume}{B595}},
  \bibinfo{pages}{408} (\bibinfo{year}{2004}), \eprint{hep-lat/0405002}.

\bibitem[{\citenamefont{Guadagnoli et~al.}(2006)\citenamefont{Guadagnoli,
  Mescia, and Simula}}]{etmctwist}
\bibinfo{author}{\bibfnamefont{D.}~\bibnamefont{Guadagnoli}},
  \bibinfo{author}{\bibfnamefont{F.}~\bibnamefont{Mescia}}, \bibnamefont{and}
  \bibinfo{author}{\bibfnamefont{S.}~\bibnamefont{Simula}},
  \bibinfo{journal}{Phys.Rev.} \textbf{\bibinfo{volume}{D73}},
  \bibinfo{pages}{114504} (\bibinfo{year}{2006}), \eprint{hep-lat/0512020}.

\bibitem[{\citenamefont{Lepage et~al.}(2002)}]{gplbayes}
\bibinfo{author}{\bibfnamefont{G.~P.} \bibnamefont{Lepage}}
  \bibnamefont{et~al.}, \bibinfo{journal}{Nucl. Phys. Proc. Suppl.}
  \textbf{\bibinfo{volume}{106}}, \bibinfo{pages}{12} (\bibinfo{year}{2002}),
  \eprint{hep-lat/0110175}.

\bibitem[{\citenamefont{Beringer et~al.}(2012)}]{pdg}
\bibinfo{author}{\bibfnamefont{J.}~\bibnamefont{Beringer}}
  \bibnamefont{et~al.}, \bibinfo{journal}{Phys. Rev. D}
  \textbf{\bibinfo{volume}{86}}, \bibinfo{pages}{010001}
  (\bibinfo{year}{2012}).

\bibitem[{\citenamefont{Dudek et~al.}(2011)\citenamefont{Dudek, Edwards, Joo,
  Peardon, Richards et~al.}}]{Dudek:2011tt}
\bibinfo{author}{\bibfnamefont{J.~J.} \bibnamefont{Dudek}},
  \bibinfo{author}{\bibfnamefont{R.~G.} \bibnamefont{Edwards}},
  \bibinfo{author}{\bibfnamefont{B.}~\bibnamefont{Joo}},
  \bibinfo{author}{\bibfnamefont{M.~J.} \bibnamefont{Peardon}},
  \bibinfo{author}{\bibfnamefont{D.~G.} \bibnamefont{Richards}},
  \bibnamefont{et~al.}, \bibinfo{journal}{Phys.Rev.}
  \textbf{\bibinfo{volume}{D83}}, \bibinfo{pages}{111502}
  (\bibinfo{year}{2011}), \eprint{1102.4299}.

\bibitem[{\citenamefont{Close}(1979)}]{close}
\bibinfo{author}{\bibfnamefont{F.}~\bibnamefont{Close}},
  \emph{\bibinfo{title}{{An Introduction to Quarks and Partons}}}
  (\bibinfo{publisher}{Academic Press}, \bibinfo{year}{1979}).

\bibitem[{\citenamefont{Arnesen et~al.}(2005)\citenamefont{Arnesen, Grinstein,
  Rothstein, and Stewart}}]{arnesenzexp}
\bibinfo{author}{\bibfnamefont{M.~C.} \bibnamefont{Arnesen}},
  \bibinfo{author}{\bibfnamefont{B.}~\bibnamefont{Grinstein}},
  \bibinfo{author}{\bibfnamefont{I.~Z.} \bibnamefont{Rothstein}},
  \bibnamefont{and} \bibinfo{author}{\bibfnamefont{I.~W.}
  \bibnamefont{Stewart}}, \bibinfo{journal}{Phys.Rev.Lett.}
  \textbf{\bibinfo{volume}{95}}, \bibinfo{pages}{071802}
  (\bibinfo{year}{2005}), \eprint{hep-ph/0504209}.

\bibitem[{\citenamefont{Hill}(2007)}]{hillzexp}
\bibinfo{author}{\bibfnamefont{R.~J.} \bibnamefont{Hill}},
  \bibinfo{journal}{eConf} \textbf{\bibinfo{volume}{C070805}},
  \bibinfo{pages}{22} (\bibinfo{year}{2007}), \eprint{0712.3817}.

\bibitem[{\citenamefont{Bourrely et~al.}(2009)\citenamefont{Bourrely, Caprini,
  and Lellouch}}]{lellouchzexp}
\bibinfo{author}{\bibfnamefont{C.}~\bibnamefont{Bourrely}},
  \bibinfo{author}{\bibfnamefont{I.}~\bibnamefont{Caprini}}, \bibnamefont{and}
  \bibinfo{author}{\bibfnamefont{L.}~\bibnamefont{Lellouch}},
  \bibinfo{journal}{Phys.Rev.} \textbf{\bibinfo{volume}{D79}},
  \bibinfo{pages}{013008} (\bibinfo{year}{2009}), \eprint{0807.2722}.

\bibitem[{\citenamefont{Na et~al.}(2010)\citenamefont{Na, Davies, Follana,
  Lepage, and Shigemitsu}}]{Na:2010uf}
\bibinfo{author}{\bibfnamefont{H.}~\bibnamefont{Na}},
  \bibinfo{author}{\bibfnamefont{C.~T.} \bibnamefont{Davies}},
  \bibinfo{author}{\bibfnamefont{E.}~\bibnamefont{Follana}},
  \bibinfo{author}{\bibfnamefont{G.~P.} \bibnamefont{Lepage}},
  \bibnamefont{and}
  \bibinfo{author}{\bibfnamefont{J.}~\bibnamefont{Shigemitsu}}
  (\bibinfo{collaboration}{HPQCD Collaboration}), \bibinfo{journal}{Phys.Rev.}
  \textbf{\bibinfo{volume}{D82}}, \bibinfo{pages}{114506}
  (\bibinfo{year}{2010}), \eprint{1008.4562}.

\bibitem[{\citenamefont{Link et~al.}(2004)}]{Link:2004qt}
\bibinfo{author}{\bibfnamefont{J.}~\bibnamefont{Link}} \bibnamefont{et~al.}
  (\bibinfo{collaboration}{FOCUS Collaboration}), \bibinfo{journal}{Phys.Lett.}
  \textbf{\bibinfo{volume}{B586}}, \bibinfo{pages}{183} (\bibinfo{year}{2004}),
  \eprint{hep-ex/0401001}.

\bibitem[{\citenamefont{Chakraborty et~al.}(2013)}]{bipasha}
\bibinfo{author}{\bibfnamefont{B.}~\bibnamefont{Chakraborty}}
  \bibnamefont{et~al.} (\bibinfo{collaboration}{HPQCD Collaboration}),
  \bibinfo{journal}{PoS} \textbf{\bibinfo{volume}{LATTICE2013}}
  (\bibinfo{year}{2013}).

\bibitem[{\citenamefont{McNeile et~al.}(2010)\citenamefont{McNeile, Davies,
  Follana, Hornbostel, and Lepage}}]{McNeile:2010ji}
\bibinfo{author}{\bibfnamefont{C.}~\bibnamefont{McNeile}},
  \bibinfo{author}{\bibfnamefont{C.}~\bibnamefont{Davies}},
  \bibinfo{author}{\bibfnamefont{E.}~\bibnamefont{Follana}},
  \bibinfo{author}{\bibfnamefont{K.}~\bibnamefont{Hornbostel}},
  \bibnamefont{and} \bibinfo{author}{\bibfnamefont{G.}~\bibnamefont{Lepage}}
  (\bibinfo{collaboration}{HPQCD Collaboration}), \bibinfo{journal}{Phys.Rev.}
  \textbf{\bibinfo{volume}{D82}}, \bibinfo{pages}{034512}
  (\bibinfo{year}{2010}), \eprint{1004.4285}.

\bibitem[{\citenamefont{McNeile
  et~al.}(2012{\natexlab{a}})\citenamefont{McNeile, Davies, Follana,
  Hornbostel, and Lepage}}]{McNeile:2011ng}
\bibinfo{author}{\bibfnamefont{C.}~\bibnamefont{McNeile}},
  \bibinfo{author}{\bibfnamefont{C.}~\bibnamefont{Davies}},
  \bibinfo{author}{\bibfnamefont{E.}~\bibnamefont{Follana}},
  \bibinfo{author}{\bibfnamefont{K.}~\bibnamefont{Hornbostel}},
  \bibnamefont{and} \bibinfo{author}{\bibfnamefont{G.}~\bibnamefont{Lepage}}
  (\bibinfo{collaboration}{HPQCD Collaboration}), \bibinfo{journal}{Phys.Rev.}
  \textbf{\bibinfo{volume}{D85}}, \bibinfo{pages}{031503}
  (\bibinfo{year}{2012}{\natexlab{a}}), \eprint{1110.4510}.

\bibitem[{\citenamefont{McNeile
  et~al.}(2012{\natexlab{b}})\citenamefont{McNeile, Davies, Follana,
  Hornbostel, and Lepage}}]{McNeile:2012qf}
\bibinfo{author}{\bibfnamefont{C.}~\bibnamefont{McNeile}},
  \bibinfo{author}{\bibfnamefont{C.}~\bibnamefont{Davies}},
  \bibinfo{author}{\bibfnamefont{E.}~\bibnamefont{Follana}},
  \bibinfo{author}{\bibfnamefont{K.}~\bibnamefont{Hornbostel}},
  \bibnamefont{and} \bibinfo{author}{\bibfnamefont{G.}~\bibnamefont{Lepage}}
  (\bibinfo{collaboration}{HPQCD Collaboration}), \bibinfo{journal}{Phys.Rev.}
  \textbf{\bibinfo{volume}{D86}}, \bibinfo{pages}{074503}
  (\bibinfo{year}{2012}{\natexlab{b}}), \eprint{1207.0994}.

\bibitem[{\citenamefont{Koponen et~al.}(2011)}]{jklatt11}
\bibinfo{author}{\bibfnamefont{J.}~\bibnamefont{Koponen}} \bibnamefont{et~al.}
  (\bibinfo{collaboration}{HPQCD Collaboration}), \bibinfo{journal}{PoS}
  \textbf{\bibinfo{volume}{LATTICE2011}}, \bibinfo{pages}{286}
  (\bibinfo{year}{2011}), \eprint{1111.0225}.

\end{thebibliography}

\end{document}